\documentclass[conference]{IEEEtran}

\ifCLASSINFOpdf
\else
\fi
\usepackage{amsmath}
\usepackage[nameinlink,capitalise]{cleveref}

\usepackage{algorithm}
\usepackage{algpseudocode}

\usepackage{amsfonts}
\usepackage{microtype}
\usepackage{hyphenat}
\usepackage{ragged2e}
\usepackage{float}
\usepackage[table]{xcolor}
\usepackage{diagbox} 
\usepackage{multirow} 
\usepackage{makecell}
\usepackage{array} 
\usepackage{graphicx}
\usepackage{subcaption}
\usepackage{tabularx}
\usepackage{booktabs}      
\usepackage{multirow}     
\usepackage{siunitx}

\begin{document}
%

\title{From Compression to Accountability: Harmless Copyright Protection for Dataset Distillation}

	

%
\author{\IEEEauthorblockN{Yan Liang\IEEEauthorrefmark{1},
Ziyuan Yang\IEEEauthorrefmark{2}, 
Mengyu Sun\IEEEauthorrefmark{1},
Joey Tianyi Zhou\IEEEauthorrefmark{3} and
Yi Zhang\IEEEauthorrefmark{1}}
\IEEEauthorblockA{\IEEEauthorrefmark{1}Sichuan University}
\IEEEauthorblockA{\IEEEauthorrefmark{2}Nanyang Technological University}
\IEEEauthorblockA{\IEEEauthorrefmark{3}Agency for Science, Technology and Research (A*STAR)}}


\IEEEoverridecommandlockouts
\makeatletter\def\@IEEEpubidpullup{6.5\baselineskip}\makeatother
\IEEEpubid{\parbox{\columnwidth}{
		Network and Distributed System Security (NDSS) Symposium 2026\\
		23 - 27 February 2026 , San Diego, CA, USA\\
		ISBN 979-8-9919276-8-0\\  
		https://dx.doi.org/10.14722/ndss.2026.[23$|$24]xxxx\\
		www.ndss-symposium.org
}
\hspace{\columnsep}\makebox[\columnwidth]{}}

\maketitle

\begin{abstract}
Large-scale datasets have been a key driving force behind the rapid progress of deep learning, but their storage, computational, and energy costs have become increasingly prohibitive. Dataset distillation (DD) mitigates this problem by synthesizing compact yet informative datasets, thereby enabling efficient model training and storage. However, the ease of copying and distributing distilled datasets introduces serious risks of copyright infringement and data leakage. Existing protection methods are primarily designed for raw datasets rather than distilled datasets, and typically rely on backdoor-triggered malicious behaviors, which may raise security concerns. In this paper, we observe that deep neural networks tend to memorize subpopulation distributions during training, resulting in a systematic prediction bias, where models perform better on samples aligned with memorized subpopulations. Motivated by this observation, we propose \textit{SubPopMark}, a harmless subpopulation-driven protection framework for distilled datasets.
SubPopMark consists of two stages. First, the Copyright Verification Marker~(CVM) optimization stage injects a class‑consistent subpopulation bias while preserving the original optimization trajectory. Second, the User‑Specific Tracing Marker (USTM) optimization stage further introduces user‑distinguishable perturbations into the CVM‑augmented data. To enable black-box verification and tracing, we construct a reference behavior bank by collecting model outputs over carefully designed test sets that cover both standard and subpopulation-shifted data distributions. The provenance of a suspicious model is then inferred by comparing its output behavior signature with the bank and identifying the most consistent reference behavior pattern. Notably, both markers avoid malicious or backdoor behavior; instead, they only bias models toward improved performance on specific subpopulations. Importantly, our method operates in a fully asymmetric setting, where we neither modify the upstream distillation process nor impose any constraints on downstream training, making the protection problem particularly challenging.
Extensive experiments demonstrate that SubPopMark enables effective copyright verification and data leakage tracing while preserving the utility of distilled datasets across various DD methods, datasets, and training protocols. In some cases, protection can be completed within 10 seconds.\footnote{The code will be made publicly available; it is omitted here for anonymity.}
\end{abstract}


%
\IEEEpeerreviewmaketitle

\section{Introduction}
\label{sec:intro}
In recent years, deep learning (DL) has achieved remarkable success, driven by rapid advances in computational resources and the availability of large-scale datasets~\cite{lei2023comprehensive}. This success is largely attributed to deep neural networks (DNNs), which can learn powerful representations from massive data and have become foundational to a wide range of applications. In both academia and industry, training on millions of samples is now the standard for achieving promising performance~\cite{devlin2019bert,hu2020randla,rong2020self}.
However, this reliance on large-scale data also leads to substantial resource consumption, including increased storage demands, prolonged training cycle, and significant energy consumption~\cite{schwartz2020green,liang2026trustunreliabilityinwardbackward}.

To address these challenges, dataset distillation (DD) has recently emerged as a promising approach~\cite{du2024diversity}. It aims to synthesize a compact yet informative dataset that preserves the essential knowledge of the raw dataset, which enables efficient training with significantly reduced storage, computational, and energy costs~\cite{sun2024diversity}. Beyond efficiency, models trained on distilled datasets can often achieve performance comparable to those trained on the full dataset~\cite{zhao2023dataset}. Owing to these advantages, DD is increasingly recognized as a practical solution for data sharing and is expected to play an important role in a wide range of machine learning applications~\cite{yu2024teddy}.


Distilled datasets are compact yet highly informative, making them particularly attractive for efficient model training while also easy to distribute and replicate~\cite{yang2025dark}. However, this unique property raises practical concerns in data-sharing scenarios, where once released, distilled data can be easily copied, redistributed, or incorporated into unauthorized applications.
This motivates the need for protection techniques that enable copyright protection and leakage tracing while preserving the utility of the distilled data.

Despite this need, research on protecting distilled datasets remains limited. Most existing protection methods are designed for raw datasets, where ownership is typically established by embedding watermarks into the training data and verifying whether the resulting model exhibits predefined behaviors associated with the watermark. However, such approaches rely on two key assumptions: (1) the training data can be directly modified, and (2) the training and inference data follow the same distribution.

These assumptions do not hold in the context of dataset distillation. From an information-theoretic perspective, compressing a large number of raw samples into a much smaller set of synthetic data inevitably results in substantial information loss, making it unrealistic for distilled datasets to faithfully preserve the raw data distribution. More importantly, dataset distillation focuses on matching optimization trajectories rather than data distributions, such that models trained on distilled data mimic the training dynamics of those trained on the full dataset. As a result, distilled datasets are inherently abstract representations that do not explicitly retain raw data instances or their distributions.

Recent studies have explored backdoor-based attacks on distilled datasets without requiring access to the raw data. While such methods can potentially be adapted for dataset copyright protection, they fundamentally rely on adversarial assumptions and introduce malicious behaviors during data construction or training. Moreover, they lack reliable traceability, making it difficult to identify or attribute the source of data leakage. Hence, we aim to address the following challenging yet important question:

\textbf{\textit{``How can we protect the copyright of distilled datasets and enable reliable traceability without relying on malicious backdoor mechanisms?"}}
\begin{figure}[!t]
    \centering
    \includegraphics[width=\columnwidth]{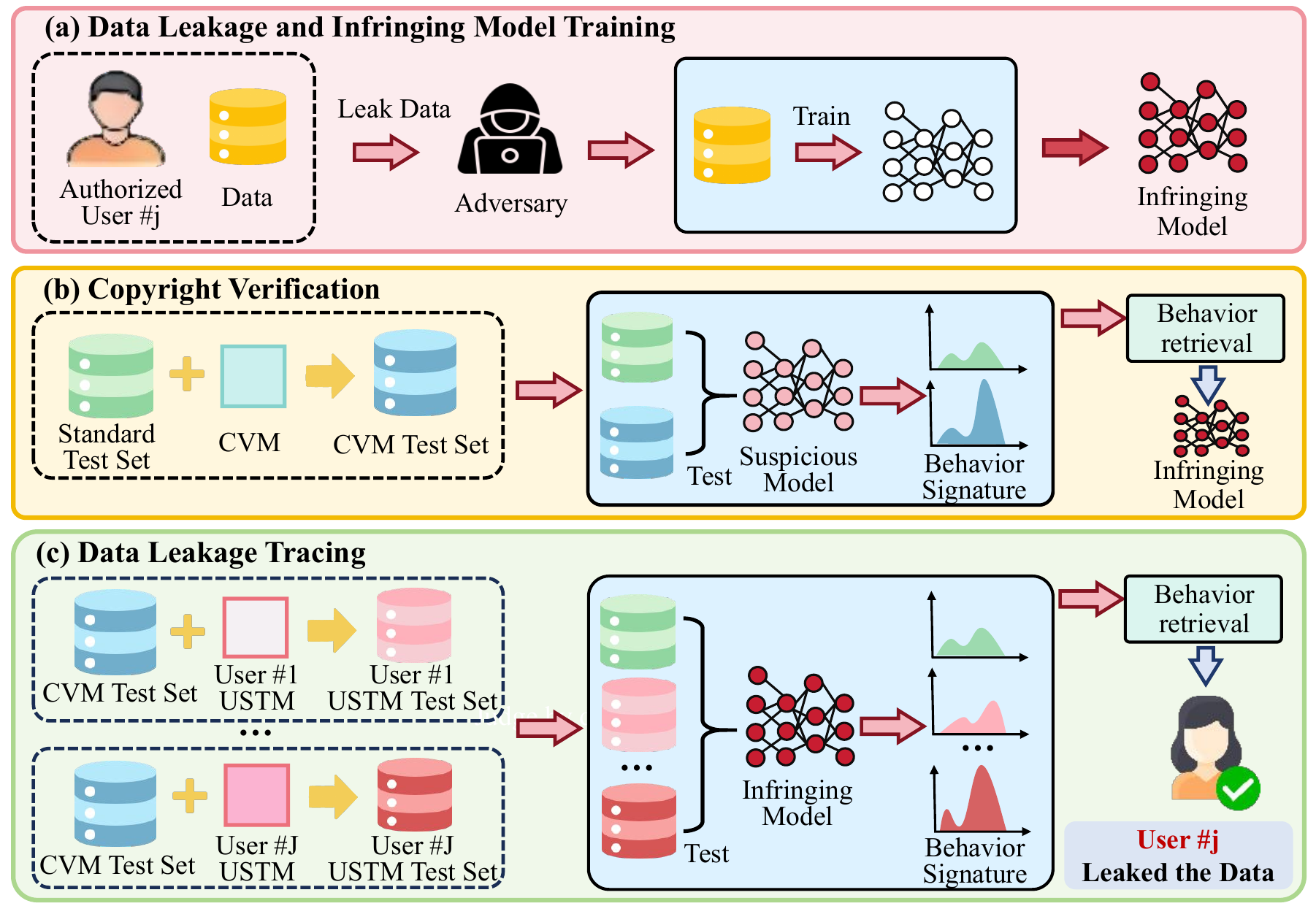}
    \caption{(a) Data Leakage Scenario: An authorized user redistributes protected distilled data to an adversary for the training of an infringing model. (b) Copyright Verification: Identification of whether a suspicious model is infringing. (c) Data Leakage Tracing: Locate the user responsible for data leakage.}
    \vspace{-10pt}
    \label{fig:intro}
\end{figure}

We further consider a realistic post-distillation protection setting. Since DD is computationally expensive, it is often impractical to re-run the distillation procedure for protection purposes. Moreover, in many practical scenarios, the data owner and the distillation provider are different parties, making it difficult to ensure that protection mechanisms are incorporated during distillation. In addition, the distillation pipeline itself is often proprietary and inaccessible to the copyright owner. Therefore, we assume that no information about the raw dataset, the DD method, or the downstream training configuration is available, and the protection mechanism must be designed without relying on any of these components. This assumption leads to a \textit{asymmetric} setting, where the defender (i.e., the distilled dataset owner) only has access to the released distilled dataset, with no visibility into the upstream DD or the downstream training pipeline.

To alleviate these issues, we observe that deep neural networks exhibit non-uniform generalization behavior even within the same class. In practice, samples from a single class often come from multiple latent subpopulations, and models tend to remember the dominant or frequently observed ones during training. As a result, learned representations are influenced not only by class semantics but also by the underlying within-class data distribution. This observation suggests that model behavior can be subtly controlled by modifying subpopulation structures without changing class labels, which provides a potential avenue for protection.

Specifically, models trained with a specific subpopulation tend to perform better on similar samples at inference time. In contrast, models trained without such subpopulation data treat these samples as distributional shifts and exhibit degraded performance. This leads to a measurable and reproducible performance gap that reflects whether a model has been trained on a certain data distribution. Importantly, this effect arises naturally from within-class distributional bias in representation learning, rather than any form of label manipulation or adversarial supervision. Therefore, it provides a benign mechanism for embedding markers into the training process, enabling copyright protection and traceability through subpopulation-induced behavioral signatures.

Based on this insight, we propose \textbf{\textit{Sub}}\textbf{\textit{Pop}}ulation-based \textit{\textbf{Mark}} (\textit{\textbf{SubPopMark}}), a harmless framework for copyright verification and data leakage tracing of distilled datasets. SubPopMark adopts a two-stage design that leverages subpopulation-induced sensitivity and a behavior signature retrieval strategy to achieve both objectives, as illustrated in Figure~\ref{fig:intro}.
Specifically, our method consists of two stages: the \textit{Copyright Verification Marker~(CVM)} optimization stage and the \textit{User-Specific Tracing Marker~(USTM)} optimization stage. In the first stage, we optimize a CVM that is integrated with the distilled dataset to inject a class-consistent subpopulation bias while preserving the original optimization trajectory. This produces a protected dataset that maintains model utility while enabling trained models to encode identifiable subpopulation-aware representations. In the second stage, we build upon the CVM to learn a USTM, which introduces user-specific subpopulation perturbations. This allows downstream models to retain copyright verification capability while additionally capturing user-distinguishable behaviors for fine-grained leakage tracing. At inference time, we evaluate a suspicious model on test sets spanning both standard and subpopulation-shifted distributions to obtain its behavior signature. This signature is then matched against a reference behavior bank constructed from models with known provenance, enabling both infringement detection and source attribution. Notably, our method operates in a fully asymmetric setting, where the defender has no access to the raw data, the distillation process, or the downstream training pipeline. The main contributions of this paper are as follows:
\begin{enumerate}
\item We study a novel post-distillation, asymmetric copyright protection setting, where the defender only has access to the released distilled dataset, without any knowledge of the raw data, distillation process, or downstream training pipeline.
\item We propose SubPopMark, the first harmless and traceable protection framework for distilled datasets, which embeds CVM and USTM to enable both copyright verification and fine-grained data leakage tracing.
\item We leverage subpopulation-induced representation bias and introduce a behavior signature retrieval, enabling downstream-agnostic copyright verification and reliable data leakage tracing without introducing backdoor or adversarial behaviors.
\end{enumerate}

\section{Related Work}
\label{sec:Related Work}

\subsection{Dataset Distillation}
\label{subsec:Dataset Distillation}
DD aims to compress a large-scale dataset {$\mathcal{D}_{\text{raw}}$} into a smaller synthetic dataset $\mathcal{D}_{\text{syn}}$ ($|\mathcal{D}_{\text{raw}}|\ll\mathcal{D}_{\text{syn}}$), with models trained on the two datasets achieving comparable performance~\cite{wang2018dataset,liu2023backdoor}.
For example, Wang \textit{et al.}~\cite{wang2018dataset} formulated DD as a bi-level meta-learning framework, and optimized the synthetic dataset such that models trained on it achieve comparable performance on the raw dataset. Building upon this foundation, subsequent studies have focused on improving efficiency and scalability~\cite{loo2022efficient, chen2024provable}. Moreover, Zhao \textit{et al.}~\cite{zhao2021datasetg} introduced gradient matching to minimize the discrepancy between the gradients of models trained on distilled data and those trained on real data. Then, this method was extended to distribution matching to further improve efficiency~\cite{zhao2023dataset}. 
In addition to gradient matching and distribution matching, several other lines of DD methods have been proposed.
For example, Differentiable Siamese Augmentation (DSA)~\cite{zhao2021dataset} improves data diversity through differentiable augmentation, while patch-based image and soft label reconstruction enrich the representation capacity of synthetic samples~\cite{sun2024diversity}. More recently, trajectory-level alignment methods have been explored. The core idea of trajectory matching is to match the evolution trajectories of the model parameters during the training process~\cite{11417909}. Specifically, Cazenavette \textit{et al.}~\cite{cazenavette2022dataset} proposed Matching Training Trajectories~(MTT), which aligns the optimization trajectories of models trained on real and synthetic data and achieves similar performance. TESLA~\cite{cui2023scaling} further improves MTT by significantly reducing computational consumption, making it a more memory-efficient variant.

\subsection{Dataset Copyright Protection}
\label{subsec:Dataset Copyright Protection}
Traditional data protection techniques are primarily designed for safeguarding private data rather than openly shared or publicly distributed datasets, and most of these methods require direct access to the raw dataset. These approaches can be broadly categorized into three types. Digital watermarking, which incorporates distinguishable ownership patterns into data to enable later identification of the data owner~\cite{guo2018halftone,kadian2021robust}. Encryption, which enforces access control by encoding data and incorporating a key-based mechanism~\cite{deng2020identity,li2021visual}. Differential privacy, which perturbs data or outputs with carefully calibrated noise to ensure that individual samples cannot be reliably distinguished, thereby protecting membership information~\cite{bai2022multinomial}. These conventional approaches often struggle to balance accessibility with protection in the context of publicly shared resources. To enable protection for open datasets, Backdoor Watermark-based Dataset Ownership Verification (BW-DOV)~\cite{li2023black,tang2023did} has emerged as an effective paradigm. The key idea is to embed specially designed watermark patterns into the training data, such that models trained on the protected dataset implicitly learn corresponding backdoor behaviors. Copyright can then be verified by probing suspicious models and checking for the presence of these abnormal responses. A representative work, DVBW~\cite{li2023black}, is the first to introduce backdoor watermarking at the dataset level and validate ownership through backdoor-triggered behaviors. More recently, backdoor-based attacks on dataset distillation have also been proposed~\cite{yang2025dark,liu2023backdoor}. While these methods can be adapted for copyright protection, these methods typically rely on modifying target labels or inducing misclassification behaviors, which leads to unstable predictions and introduces potential adversarial vulnerabilities.

\section{Threat Model}
\label{sec:Threat Model}
We consider a practical post-distillation copyright protection and leakage tracing scenario. This setting is motivated by two key factors. First, DD is computationally expensive, making it impractical to re-run the distillation procedure or re-distill the dataset for protection purposes.
Second, the data owner and the distillation provider are different parties, making it difficult for the defender to ensure that the protection mechanism is properly incorporated during distillation. In addition, the distillation pipeline itself is often proprietary intellectual property and therefore inaccessible to the copyright owner.
For clarity in the following discussion, we refer to the \textit{defender} as the copyright owner of the distilled dataset who seeks to protect its copyright and trace potential data leakage.


\noindent \textbf{Adversary.} The adversary can obtain the distilled dataset through unauthorized redistribution, and subsequently use it to train their own models or further distribute the dataset to other parties without authorization. The adversary has full control over the adversary-side model, including its architecture, parameters, and training configurations.

\noindent \textbf{Defender Capability.} The defender aims to enable reliable copyright verification and accurate leakage tracing over released distilled datasets in a strictly constrained and partially observable setting. Overall, the defender operates under the following constraints:

\noindent \textit{Restricted-Access.} The defender operates in a restricted-access asymmetric setting, where there is no access to the upstream dataset or distillation process, nor any visibility into the downstream training pipeline or internal parameters of models trained on potentially leaked data.



\noindent \textbf{Defender Goal.} The defender aims to establish effective copyright protection mechanisms for released distilled datasets, enabling both reliable copyright verification and precise leakage tracing. Overall, the defender is required to achieve the following objectives:

\noindent \textit{Copyright Verification.} Determine whether a suspicious model has been trained on the protected distilled dataset, without requiring access to model parameters or training details, and under the assumption that only query-based interactions with the model are available.

\noindent \textit{Leakage Tracing.} Identify the specific authorized user responsible for unauthorized dataset redistribution when infringement is detected, by distinguishing user-specific test set variants and attributing the observed model behavior to the corresponding data source.

\noindent \textit{Benign Design.} Ensure that the protection mechanism does not rely on any adversarial or malicious behaviors, and achieves verification and tracing solely through the benign behavior signatures.

\noindent \textbf{Challenges.} We summarize two key challenges in designing effective post-distillation copyright protection and leakage tracing mechanisms.

\noindent  \textbf{(i) The Gap between the distilled data and the real data.} Unlike raw datasets, distilled datasets are highly compressed and distributionally abstract representations, which discard instance-level correspondence and fine-grained statistical structure. As a result, the defender cannot rely on explicit sample-level markers or reconstruction-based cues, making it difficult to embed reliable and verifiable ownership signals without altering the learning objective or degrading data utility.

\noindent \textbf{(ii) Robust Attribution under Asymmetric and Dynamic Training Environments.}
Beyond the asymmetric access constraint, the attribution signal must remain reliable in dynamic training environments. The protection mechanism is not allowed to rely on explicit supervision or trigger-based injections~(e.g., backdoors)to preserve benignness and dataset utility. Therefore, the key challenge is to embed an implicit yet robust dataset-level structure that consistently induces distinguishable model behaviors across diverse training conditions while remaining imperceptible and non-invasive.

\section{Theory Analysis}
\label{sec:theory}
The core principle of our framework is that neural networks naturally assign higher confidence to inputs that follow their training data distribution, and lower confidence to inputs that lie outside that distribution~\cite{Lei_2026_ZePAD}. In other words, the model exhibits a consistent behavioral signature toward inputs aligned with its training distribution. This behavior is not manually designed, but emerges naturally from empirical risk minimization using cross-entropy in training. In this section, we present a theoretical analysis to support our method.

\subsection{Definition}
Let $\mathcal{D}_{\text{syn}}=\{(\tilde{x},\tilde{y})\}$ denote the distilled dataset, where $\tilde{x}$ is the distilled image and $\tilde{y}$ is the corresponding class label. Let $\mathcal{M}$ denote a subpopulation transformation operator. We define $\mathcal{D}_{\mathcal{M}}$ as the transformed distribution induced by applying $\mathcal{M}$ to samples from $\mathcal{D}_{\text{syn}}$, i.e., $(\mathcal{M}(\tilde{x}), \tilde{y}) \sim \mathcal{D}_{\mathcal{M}}$ for $(\tilde{x},\tilde{y}) \sim \mathcal{D}_{\text{syn}}$.
Then, the protected distilled data distribution can be formulated as follows:
\begin{equation}
\small
\mathcal{D}_{\text{mark}}
=
\left(
\mathcal{D}_{\text{syn}} \setminus \mathcal{S}
\right)
\cup 
\mathcal{D}_{\mathcal{M}},
\mathcal{D}_{\mathcal{M}}
=
\left\{
(\mathcal{M}(\tilde{x}),\tilde{y})
\;\middle|\;
(\tilde{x},\tilde{y}) \in \mathcal{S}
\right\},
\end{equation}
where $|\mathcal{S}|=\alpha |\mathcal{D}_{\text{syn}}|$. $\alpha$ is the sampling probability, and $S$ denotes the manipulation subset.

Then, we define that the reference model $\theta_R$ and the infringing model $\theta_I$ are trained on $\mathcal{D}_{\text{syn}}$ and $\mathcal{D}_{\text{mark}}$, respectively.

\subsection{Information Theory Analysis}
Consider a model $\theta$ trained using the cross-entropy loss:
\begin{equation}
    \mathcal{L}(\theta) = \mathbb{E}_{(\tilde{x},\tilde{y})\sim \mathcal{D}_{\text{syn}}}[-\log \theta(\tilde{x})_{\tilde{y}}],
\end{equation}
where $\theta_{\tilde{y}}(\tilde{x})$ denotes the predicted probability for class $\tilde{y}$.

Under sufficient model capacity and training convergence, the learned classifier approximates the true posterior distribution~\cite{Lei_2026_ZePAD,zhang2004statistical}:
\begin{equation}
\small
    \theta^\star(\tilde{x}) \approx P(\tilde{y} \mid \tilde{x}),
\end{equation}
where $\theta^\star$ denotes the theoretically optimal model. $P$ denotes the real data distribution.

For our reference and infringing models, we have $\theta_R(\tilde{x}) \approx P_{\text{syn}}(\tilde{y}\mid\tilde{x})$ and $\theta_I(\tilde{x}) \approx P_{\text{mark}}(\tilde{y}\mid\tilde{x})$, where $P_{\text{syn}}$ and $P_{\text{mark}}$ denote the original distilled dataset and the protected distilled dataset distributions, respectively. $\theta_R(\tilde{x})$ and $\theta_I(\tilde{x})$ denote the output probability vectors of $\theta_R$ and $\theta_I$, respectively. Then, we define the model prediction confidence using the maximum softmax probability:
\begin{align}
    m_{R}(\tilde{x}) = \max_k \theta_{R}(\tilde{x})_k \approx P_{\text{syn}}(k \mid \tilde{x}),\\
    m_{I}(\tilde{x}) = \max_k \theta_{I}(\tilde{x})_k \approx P_{\text{mark}}(k \mid \tilde{x}),
\end{align}
where $m_{\text{R}}(\tilde{x})$ and $m_{\text{I}}(\tilde{x})$ denote the confidence scores of the reference and infringing models, respectively, reflecting the model predictions induced by their respective training data distributions. Specifically, $\theta_{\text{R}}(\tilde{x})_k$ (or $\theta_{\text{I}}(\tilde{x})_k$) denotes the predicted probability that the input $\tilde{x}$ belongs to class $k$.

\subsection{Distribution-Induced Behavioral Bias}

\begin{figure*}[htbp]
    \centering
    \includegraphics[width=\textwidth]{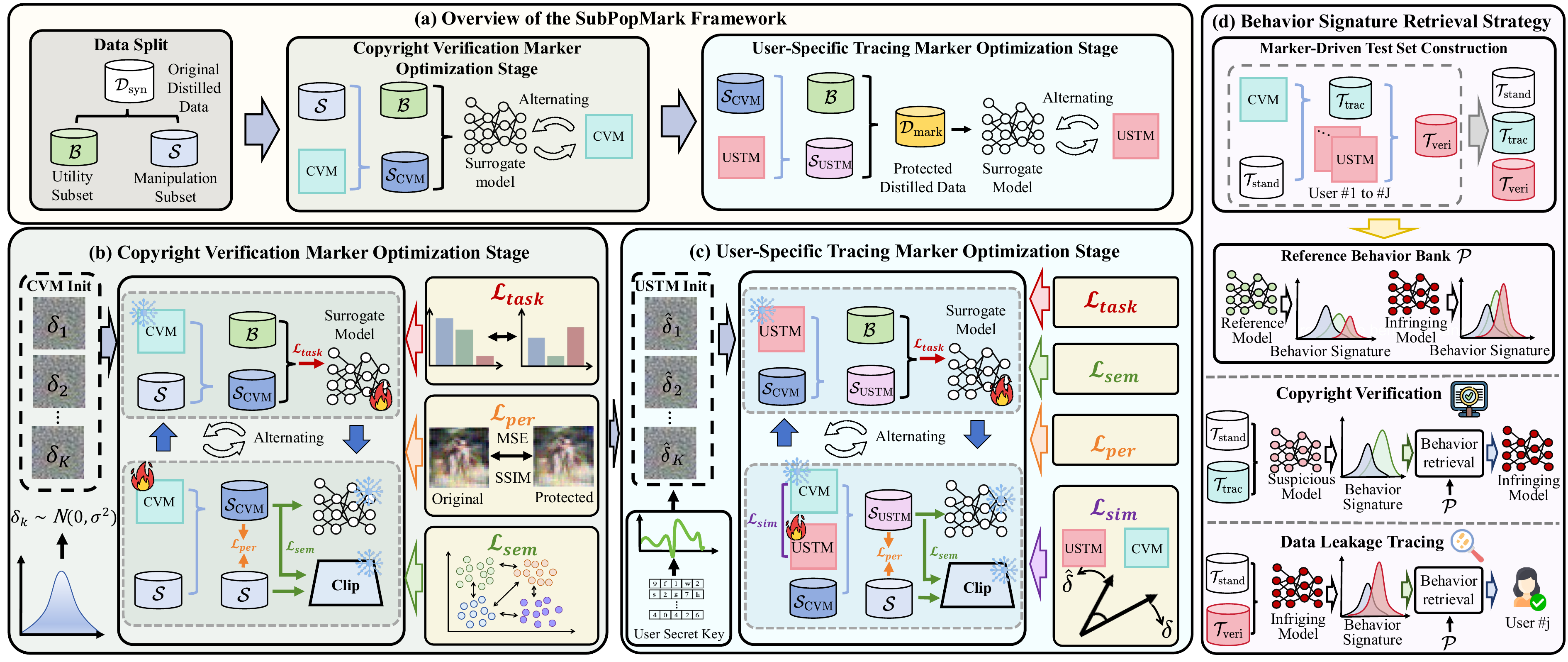}
    \caption{The overview of our proposed method.}
    \vspace{-20pt}
    \label{fig:framework}
\end{figure*}

For samples $(x,y)\sim\mathcal{T}$, where $\mathcal{T}$ denotes the test data distribution, due to the objectives and properties of the distilled dataset. As discussed earlier, the posterior distribution of $\theta_I$ is governed by $\mathcal{D}_\text{mark}$. Since subpopulation markers preserve semantic labels and the injection ratio $\alpha$ is small, the posterior satisfies:
\begin{equation}
    P_{\text{mark}}(y\mid x) \approx P_{{\text{syn}}}(y\mid x),
\end{equation}
which implies:
\begin{equation}
\small
    m_{\text{I}}(x) \approx m_{\text{R}}(x).
\end{equation}

In contrast, for the sample $(x, y) \sim \mathcal{T}_{\mathcal{M}}$, where $\mathcal{T}_{\mathcal{M}}$ denotes the distribution obtained by applying transformation $\mathcal{M}$ to samples from distribution $\mathcal{T}$, the infringing model has learned the bias introduced by $\mathcal{M}$ during training, whereas the reference model has not. As a result, the infringing model better aligns with $\mathcal{D}_{\mathcal{M}}$, leading to:
\begin{equation}
\small
P_{{\text{mark}}}(y\mid x)
= (P_{\text{syn}}(y\mid x)\setminus P_{\mathcal{S}}(y\mid x))
\cup
P_{\mathcal{M}}(y\mid x),
\end{equation}
where $P_{\mathcal{M}}$ and $P_{\mathcal{M}}$ denote the manipulation subset and transformed distributions.
Since the transformed samples follow a consistent subpopulation structure introduced during training, the infringing model assigns a higher posterior probability to the correct label compared to the reference model. Therefore, we expect:
\begin{equation}
    P_{\text{mark}}(y\mid x)  > P_{\text{syn}}(y\mid x).
\end{equation}
Consequently, the confidence satisfies:
\begin{equation}
\small
    m_{\text{I}}(x) > m_{\text{R}}(x).
\end{equation}

As proved above, we can obtain that the samples form the raw test data distribution, the confidence of the reference model and the infringing model are essentially consistent. Conversely, for inputs with the SubPopMark, the infringing model yields a significantly higher confidence than the reference model. This confirms a clear confidence separation between the models, demonstrating the inherent sensitivity of neural networks to different subpopulation distributions. Specifically, a network tends to assign higher confidence to samples that align with its training distribution, while yielding lower confidence on unseen or distribution-shifted samples.

\section{Proposed Method}
\subsection{Problem Formulation}
As mentioned earlier, DD aims to extract knowledge from a large-scale dataset $\mathcal{D}_{\text{raw}}$ and construct a much smaller synthetic dataset $\mathcal{D}_{\text{syn}}$ ($|\mathcal{D}_{\text{raw}}| \gg |\mathcal{D}_{\text{syn}}|$). The goal is that the downstream model $\theta_{\mathcal{D}_{\text{syn}}}$ trained on $\mathcal{D}_{\text{syn}}$ can achieve performance comparable to that of a model $\theta_{\mathcal{D}_{\text{raw}}}$ trained on $\mathcal{D}_{\text{raw}}$, when evaluated over the raw test data distribution $\mathcal{T}$:
\begin{equation}
    \mathbb{E}_{(x,y) \sim \mathcal{T}} [\ell(\theta_{\mathcal{D}_{\text{syn}}}(x), y)] \simeq \mathbb{E}_{(x,y) \sim \mathcal{T}} [\ell(\theta_{\mathcal{D}_{\text{raw}}}(x), y)],
\end{equation}
where $\ell$ denotes the loss function between the prediction and the ground truth $y$.

In this paper, we investigate how to verify whether a deployed model is trained on unauthorized distilled datasets, while further supporting user-level data leakage tracing. We construct a protected dataset $\mathcal{D}_\text{CVM}$ based on $\mathcal{D}_{\text{syn}}$. Given a suspicious model $\theta_s$, our goal is to determine whether it has been trained on $\mathcal{D}_\text{CVM}$.
We construct a verification test set $\mathcal{T}_{\text{veri}}$ based on a standard test set $\mathcal{T}_{\text{stand}}$ to expose behavioral biases induced by $\mathcal{D}_{\text{CVM}}$, which can be formulated as:
\begin{equation}
    \mathcal{R}(\theta_{\text{CVM}}, \mathcal{T}_{\text{veri}}) \gg \mathcal{R}(\theta_{\text{syn}}, \mathcal{T}_{\text{veri}}),
\end{equation}
where $\theta_{\text{CVM}}$ and $\theta_{\text{syn}}$ denote models trained on $\mathcal{D}_{\text{CVM}}$ and $\mathcal{D}_{\text{syn}}$. $\mathcal{R}(\theta, \mathcal{T})$ denotes the classification accuracy achieved by the model $\theta$ on the test set $\mathcal{T}$,  which can be defined as:
\begin{equation}
\label{eq:acc}
\mathcal{R}(\theta, \mathcal{T}) 
= \mathbb{E}_{(x,y)\sim \mathcal{T}} 
\left[ \mathbb{I}\big(\arg\max_k \theta(x)_k = y\big) \right],
\end{equation}

We define the copyright verification performance gap as:
\begin{equation}
\small
    G(\theta) = \mathcal{R}(\theta, \mathcal{T}_{\text{stand}}) - \mathcal{R}(\theta, \mathcal{T}_{\text{veri}}),
\end{equation}
where $G(\cdot)$ measures the discrepancy between the performance on the standard and verificiation test sets. The verification decision is then formulated as a hypothesis test:
\begin{equation}
\small
\mathcal{V}(\theta_s) =
\begin{cases}
1, & \text{if } G(\theta_s) \simeq G(\theta_{\text{CVM}}) \\
0, & otherwise
\end{cases},
\end{equation}
where $\mathcal{V}(\theta_s)=1$ indicates that the suspicious model $\theta_s$ is identified as an infringing model.

Except copyright verification, we further aim to trace the user responsible for the underlying data leakage. Specifically, we add the user-specific marker in $\mathcal{D}_\text{CVM}$ to generate $\mathcal{D}_{mark}=\{\mathcal{D}^{j}_\text{mark}\}_{j=1}^J$, where $J$ is the number of users. Each dataset is constructed to preserve the same copyright verification capability as $\mathcal{D}_{\text{CVM}}$, while additionally enabling fine-grained data leakage tracing to individual users. Similar with the previous step, we can construct $\{\mathcal{T}^j_{\text{trac}}\}_{j=1}^J$ for each user. Then, we can get the user-specific leakage tracing performance gap:
\begin{equation}
\small
    \hat{G}^j(\theta) = \mathcal{R}(\theta, \mathcal{T}_{\text{stand}}) - \mathcal{R}(\theta, \mathcal{T}^j_{\text{trac}}).
\end{equation}
The estimated leaking user index $\hat{j}$ is then obtained as:
\begin{equation}
\small
\label{eq:tracing}
\hat{j} = {\arg\max}_{j \in \{1, \dots, J\}} |\hat{G}^j(\theta_I)|.
\end{equation}

\subsection{Overview of SubPopMark}
In this paper, we propose the \textit{SubPopMark} framework, which manipulates the distilled dataset to induce a subpopulation-specific behavior bias in models trained on the protected distilled dataset, while preserving their performance compared to those trained on the original distilled dataset.
Specifically, SubPopMark consists of two stages, including the \textit{Copyright Verification Marker~(CVM)} optimization stage and \textit{User-Specific Tracing Marker~(USTM)} optimization stage, and the overview of our proposed method is illustrated in Figure~\ref{fig:framework}. In the first stage, we optimize a CVM integrated into the distilled dataset to induce a class-consistent subpopulation bias while preserving the original optimization trajectory. This produces a protected dataset that enables trained models to exhibit identifiable subpopulation-aware behaviors without sacrificing utility. In the second stage, we further learn a USTM to introduce user-specific subpopulation perturbations, allowing downstream models to encode tracable behavioral signatures. During inference, we extract the behavior signature of a target model using both standard and subpopulation-shifted test sets, and match it against a reference behavior bank for infringement verification and data leakage tracing.



\subsection{Copyright Verification Marker Optimization}
\begin{algorithm}[t]
\caption{Main steps of CVM optimization}
\label{alg:cvm}
\begin{algorithmic}[1]
\Require Distilled dataset $\mathcal{D}_{\text{syn}}$, class number $K$, ratio $\alpha$, learning rates $\eta_\theta, \eta_\delta$, a frozen feature encoder $\Phi$

\Ensure Optimized CVM parameters $\delta$

\Statex \hspace*{-\algorithmicindent}\textbf{Initialize:} 
\State Initialize class-specific markers $\delta = \{\delta_k\}_{k=1}^K \sim \mathcal{N}(0, \sigma^2)$ and surrogate model $\theta$

\State Partition $\mathcal{D}_{\text{syn}}$ into:
\State \hspace{1em} manipulation subset $\mathcal{S}$ with ratio $\alpha$
\State \hspace{1em} utility subset $\mathcal{B} = \mathcal{D}_{\text{syn}} \setminus \mathcal{S}$

\For{epoch $t = 1$ to $T$}

    \State \textbf{Step 1: Update $\theta$ and fix $\delta$}
    \State Construct transformed subset:
    \[
    \mathcal{S}^{t}_{\text{CVM}} = \{(\tilde{x} + \delta^{t-1}_{\tilde{y}}, \tilde{y}) \mid (\tilde{x}, \tilde{y}) \in \mathcal{S}\}
    \]

    \State Compute $\mathcal{L}_{\text{task}}(\mathcal{B} \cup \mathcal{S}^t_{\text{CVM}}; \theta^{t-1})$ $\triangleright$ Based on Eq.~\eqref{eq:task}
    
    \State \textit{Update \textbf{$\theta$}}: $\theta^t \leftarrow \theta^t - \eta_\theta \nabla_\theta \mathcal{L}_{\text{task}}(\mathcal{B} \cup \mathcal{S}^t_{\text{CVM}}; \theta^{t-1})$

    \State \textbf{Step 2: Update $\delta$ and fix $\theta$}
    
    \State Compute $\mathcal{L}_{\text{sem}}(\mathcal{S}; \delta^{t-1})$ \hfill $\triangleright$ Based on Eq.~\eqref{eq:sem}

    \State Compute $\mathcal{L}_{\text{per}}(\mathcal{S}; \delta^{t-1})$ \hfill $\triangleright$ Based on Eq.~\eqref{eq:per}

    \State Compute $\mathcal{L}_{\text{CVM}}(\delta^{t-1}, \theta^t)$ \hfill $\triangleright$ Based on Eq.~\eqref{eq:CVM}

    \State \textit{Update \textbf{$\delta$}}: $\delta^t \leftarrow \delta^{t-1} - \eta_\delta \nabla_\delta 
    \mathcal{L}_{\text{CVM}}(\delta^{t-1}, \theta^t)$

\EndFor
\end{algorithmic}
\end{algorithm}

To enable copyright verification for a suspicious model, we introduce a CVM into the distilled dataset to construct a protected dataset, such that models trained on it acquire a verifiable behavior bias toward a subpopulation distribution.
Moreover, CVM is expected to preserve the original optimization trajectory, thereby ensuring that the resulting models maintain performance comparable to those trained on the original distilled dataset.

Specifically, the CVM is parameterized as a learnable set of class-specific markers $\delta = \{\delta_{k}\}_{k=1}^K$, where $K$ is the number of classes and $\delta_k$ matches the dimensionality of the distilled sample and is initialized from a Gaussian distribution $\mathcal{N}(0, 1)$. 
Then, the CVM-induced transformation is defined as $\mathcal{M}_{\text{CVM}}(\tilde{x}, \tilde{y}) = (\tilde{x} + \delta_{\tilde{y}}, \tilde{y})$, where $(\tilde{x}, \tilde{y})$ denotes a sample $\tilde{x}$ from $\mathcal{D}_{\text{syn}}$ with $\tilde{y}$ is the label, and $\delta_{\tilde{y}}$ is the corresponding class-specific marker. Notably, the CVM-induced transformation is applied only to the input $\tilde{x}$, while leaving $\tilde{y}$ unchanged. This design ensures that only subpopulation-level bias is introduced, as modifying labels may lead to unintended malicious behaviors.

To achieve a balance between inducing a verifiable behavior bias and preserving the performance, we partition $\mathcal{D}_{\text{syn}}$ into two disjoint subsets. We construct a manipulation subset $\mathcal{S}$ by sampling each data in $\mathcal{D}_{\text{syn}}$ with probability $\alpha$, and define the utility subset as $\mathcal{B} = \mathcal{D}_{\text{syn}} \setminus \mathcal{S}$.
Then, the CVM transformation is applied only to samples in $\mathcal{S}$, yielding the transformed subset $\mathcal{S}_{\text{CVM}} = \mathcal{M}_{\text{CVM}}(\mathcal{S})$.
The subset $\mathcal{B}$ remains unchanged.
Then, both subsets are jointly used to optimize $\delta$, where $\mathcal{S}_{\text{CVM}}$ drives the learning of the verifiable biases, and $\mathcal{B}$ ensures that the optimization remains close to the original training trajectory, thereby preserving task performance.

We first introduce a task loss to ensure that the model trained on the updated dataset remains consistent with the original task objective. The loss function can be formulated as:
\begin{equation}
\label{eq:task}
\mathcal{L}_{\text{task}}(\mathcal{B}; \theta) 
= \mathbb{E}_{(\tilde{x}, \tilde{y}) \sim (\mathcal{B} \cup\mathcal{S}_\text{CVM})} 
\left[ -\log \theta(\hat{x})_{\hat{y}} \right],
\end{equation}
where $\theta(\hat{x})_{\hat{y}}$ denotes the predicted probability for class $\hat{y}$ given input $\hat{x}$.

Then, we introduce a regularization loss $\mathcal{L}_{\text{sem}}$ to balance task utility and subpopulation distinguishability for our CVM. Specifically, we formulate the loss as:
\begin{equation}
\label{eq:sem}
\scriptsize
\mathcal{L}_{\text{sem}}(\mathcal{S}; \delta) 
= \underset{\substack{(\tilde{x}, \tilde{y}) \sim \mathcal{S} \\ \delta_{\tilde{y}} \sim \delta}}{\mathbb{E}}
\left[
- \log p_\theta(\tilde{y} \mid \tilde{x}+\delta_{\tilde{y}})
+ \mathrm{CosSim}\big(\Phi(\tilde{x}), \Phi(\tilde{x} + \delta_{\tilde{y}})\big)
\right],
\end{equation}
where $\Phi$ denotes a frozen pretrained feature encoder~\cite{radford2021learning}.
The first term enforces label consistency between the original and perturbed samples to avoid introducing any malicious backdoors. The second term ensures that the subpopulation introduced by CVM remains verifiable in the feature space, enabling reliable copyright verification.

Finally, we further constrain the perturbation to be imperceptible at the visual level, ensuring that the perturbed samples remain visually consistent with the original distilled samples. To achieve this, we introduce a perceptual loss as:
\begin{equation}
\label{eq:per}
\scriptsize
\mathcal{L}_{\text{per}}(\mathcal{S};\delta)
= \underset{\substack{(\tilde{x}, \tilde{y}) \sim \mathcal{S} \\ \delta_{\tilde{y}} \sim \delta}}{\mathbb{E}}
\Bigg[ \| \tilde{x} - (\tilde{x} + \delta_{\tilde{y}}) \|_2^2  + \bigl(1 - \mathrm{MS\text{-}SSIM}(\tilde{x},\, \tilde{x} + \delta_{\tilde{y}})\bigr)
\Bigg],
\end{equation}
where the first term penalizes pixel-level distortion to keep the perturbation small in the image space, and the second term enforces perceptual consistency via a multi-scale structural similarity (MS-SSIM)~\cite{wang2003multiscale} metric that measures visual and structural similarity between images to preserve overall appearance.

Finally, the overall loss to optimize CVM is defined as the combination of the losses introduced above:
\begin{equation}
\label{eq:CVM}
    \mathcal{L}_{\text{CVM}}(\delta, \theta) =  \mathcal{L}_{\text{sem}}(\mathcal{S};\delta) + \mathcal{L}_{\text{per}}(\mathcal{S};\delta).
\end{equation}

During optimization, we adopt an alternating training strategy to update the model parameters $\theta$ and the CVM parameters $\delta$ in an epoch-wise manner. Specifically, in each epoch, we first update $\theta$ by minimizing the task loss $\mathcal{L}_{\text{task}}$ on the combined dataset $\mathcal{B} \cup \mathcal{S}_{\text{CVM}}$, which ensures that the model follows the original training trajectory and maintains performance consistency.
In the subsequent epoch, we fix the updated $\theta$ and update the CVM $\delta$ by optimizing $\mathcal{L}_{\text{CVM}}$. This allows the CVM to adapt to the current state of the model while introducing the subpopulation representation bias and ensuring the visual imperceptibility.

Such an alternating optimization strategy aligns the optimization of CVM with the training trajectory induced by the clean distilled dataset, ensuring that the optimization remains close to the original trajectory. Importantly, this procedure does not require any access to real images, as this process is entirely conducted on the synthetic dataset. Since our protection mechanism is applied directly to the distilled dataset, whose size is typically extremely compact, the overall protection overhead remains negligible. We provide further empirical analysis of the efficiency in Sec.~\ref{sec:comp}. For clarity and reproducibility, we summarize the entire optimization procedure in \textit{Algorithm}~\ref{alg:cvm}

\subsection{User-Specific Tracing Marker Optimization}
CVM can effectively enable copyright verification by introducing a behavioral bias toward a specific subpopulation. However, since the same CVM is shared across all distributed protected datasets, the resulting behavioral bias is identical for all users, making it incapable of tracing the specific source of data leakage. To further enable reliable user-level tracing, the key idea is to associate each authorized user with a unique perturbation pattern, such that models trained on different user-specific datasets exhibit distinguishable signatures. To this end, we introduce a user-specific key for each user, which deterministically controls the generation of the corresponding markers. This design ensures both uniqueness (different users obtain different markers) and reproducibility (the same user can regenerate the marker when needed in the limited time), which are essential for reliable tracing.

The distinguishability of user-specific markers is primarily induced by their different initializations, which lead to divergent optimization trajectories. Specifically, each user-specific marker $\hat{\delta}^j$ is initialized from an independent key-controlled random process, resulting in different initial directions in the perturbation space. Under standard gradient-based optimization, the update of $\hat{\delta}^j$ can be viewed as following a trajectory conditioned on its initialization. Due to the non-convexity of the objective, different initializations typically converge to different local minima or regions in the parameter space.

Specifically, for each user $j$, we parameterize the user-specific tracing marker (USTM) as a set of class-dependent perturbations $\hat{\delta}^j = \{\hat{\delta}_k^j\}_{k=1}^K$. The markers are generated from a pseudorandom process controlled by the user-specific key. Concretely, we compute a seed $s^j = \mathcal{H}(W^j)$ from the user’s secret key $W^j$, where $\mathcal{H}$ denotes the cryptographic SHA-256 hash function~\cite{pub2012secure}, and generate each perturbation as $\hat{\delta}_k^j = \mathcal{G}(s^j, k)$, where $\mathcal{G}$ outputs Gaussian-distributed vectors. The resulting transformation is applied on top of the CVM-processed samples, i.e., $\mathcal{M}_u^j(\tilde{x}, \tilde{y}) = (\tilde{x} + \delta_{\tilde{y}} + \hat{\delta}^j_{\tilde{y}}, \tilde{y})$, thereby constructing user-specific subpopulations while preserving the original label semantics.

To optimize the USTM, we follow the same alternating training protocol as in CVM. Specifically, we first update the model parameters $\theta$ using the task loss on the user-specific dataset, ensuring that the model remains aligned with the original task objective. Then, with $\theta$ fixed, we update the user-specific markers $\hat{\delta}^j$ by optimizing a combination of semantic and perceptual objectives, allowing the markers to adapt to the current model while maintaining distinguishability and visual imperceptibility.

One consideration is to prevent user-specific markers from interfering with the shared CVM, as such interference could potentially affect the reliability of both copyright verification and data leakage tracing. To mitigate this, we introduce a feature similarity term in the feature space:
\begin{equation}
\mathcal{L}_{\text{sim}}(\hat{\delta}^j, \delta) 
= \mathbb{E}_{\hat{\delta}^j_{k} \sim \hat{\delta}^j, \delta_{k} \sim \delta} \left[ \mathrm{CosSim}\big(\Phi(\hat{\delta}_k^j), \Phi(\delta_k)\big) \right].
\end{equation}

This loss encourages user-specific markers to have low feature similarity with respect to the shared CVM among all users, aiming to keep their subpopulation biases without causing strong interference. During optimization, this term is jointly optimized with the semantic and perceptual losses when updating $\hat{\delta}^j$, while keeping the CVM training procedure unchanged.

\begin{figure*}[!t]
    \centering
    \includegraphics[width=.9\textwidth]{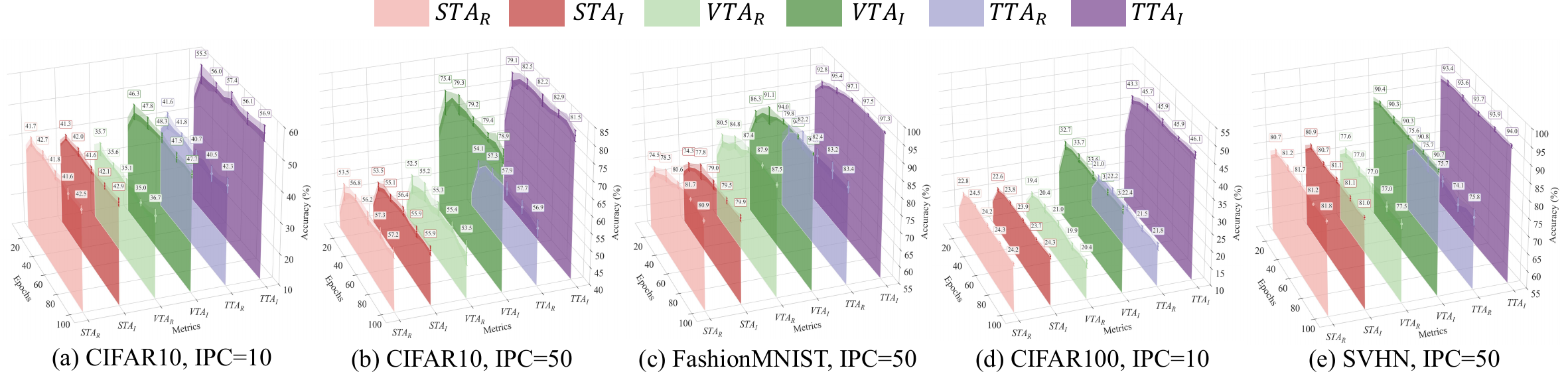}
    \caption{Performance analysis of SubPopMark across diverse datasets and IPC settings using the DC distillation method. All experimental results are averaged over 10 independent trials, with the shaded regions representing the variance.}
    \vspace{-15pt}
    \label{fig:DC_base}
\end{figure*}

\subsection{Implementation}
In our pipeline, the protected dataset $\mathcal{D}^j_{\text{mark}}$ is released to user $j$ as a replacement for the original distilled dataset $\mathcal{D}_{\text{syn}}$.
Then, each user trains their model on $\mathcal{D}^j_{\text{mark}}$ without awareness of the embedded markers. Then, each user trains their model on $\mathcal{D}^j_{\text{mark}}$ without awareness of the embedded markers. During training, the model naturally learns a bias toward the corresponding subpopulation distribution induced by the injected marks.

In practice, users typically do not release model parameters and instead expose only inference APIs. Therefore, our verification setting is inherently asymmetric, and we can only access model predictions. Moreover, our method aims to characterize the model’s behavioral preference induced by the injected perturbations, rather than detecting instance-level anomalies. To this end, we adopt a batch-based evaluation strategy, where a set of query samples is used to probe the suspicious model and estimate its preference over specific subpopulation distributions.

To facilitate such distribution-level evaluation, we construct multiple test sets that reflect the behaivor of the model to different underlying data distributions. Specifically, we define three types of test sets: a standard test set $\mathcal{T}_{\text{stand}}$ drawn from the raw test data distribution, a copyright verification test set $\mathcal{T}_{\text{veri}}$ constructed using CVM, and a collection of user-specific tracing test sets $\mathcal{T}_{\text{trac}} = \{\mathcal{T}_{\text{trac}}^j\}_{j=1}^J$ constructed using USTM.

Under this setup, an infringing model~(i.e., a model trained on the protected dataset $\mathcal{D}_{\text{mark}}^j$) develops a behavioral preference toward the corresponding marked subpopulation. As a result, compared with reference models~(i.e., models trained on the original distilled dataset $\mathcal{D}_{\text{syn}}$), infringing models exhibit systematically shifted responses on the marked test sets~(i.e., $\mathcal{T}_{\text{veri}}$ and $\mathcal{T}_{\text{trac}}^j$), while maintaining comparable behavior on the standard test set $\mathcal{T}_{\text{stand}}$. Specifically, $\mathcal{T}_{\text{veri}}$ and $\mathcal{T}_{\text{trac}}^j$ can be formulated as follows:
\begin{align}
    \mathcal{T}_{\text{veri}} = \{(x + \delta_y, y) \mid (x,y) \in \mathcal{T}_{\text{stand}}\},\\
    \mathcal{T}_{\text{trac}}^j = \{(x + \hat{\delta}_{y}^j, y) \mid (x, y) \in \mathcal{T}_{\text{veri}}\}.
\end{align}

To enable practical verification under the black-box setting, we further introduce a behavior signature retrieval strategy that operates purely on model outputs, without requiring any access to model parameters or architectural details. Specifically, we first construct a reference behavior bank by querying a collection of models with known provenance (e.g., models trained on $\mathcal{D}_{\text{syn}}$ and $\mathcal{D}_{\text{mark}}^j$) over the predefined test sets. For each model, we record its responses on $\mathcal{T}_{\text{stand}}$, $\mathcal{T}_{\text{veri}}$, and $\mathcal{T}_{\text{trac}}^j$, which can be formulated as:
\begin{equation}
\mathbf{s}(\theta) = \big[ \mathcal{R}(\theta, \mathcal{T}_{\text{stand}}),\; \mathcal{R}(\theta, \mathcal{T}_{\text{veri}}),\; \mathcal{R}(\theta, \mathcal{T}_{\text{trac}}^j) \big],
\end{equation}
where \( s(\cdot) \) denotes a function that summarizes model predictions. Then, we can construct a set of behavior signatures of the preference of different models over different data distributions. This process can be formulated as:
\begin{equation}
\mathcal{P} = \left\{ (\mathbf{s}(\theta_a), z_a) \right\}_{a=1}^A,
\end{equation}
where $A$ denotes the length of the reference behavior bank and $z_a$ indicates whether the model is an infringing model and the corresponding user identity.

We then perform behavior-based retrieval by matching the target signature against the reference behavior bank:
\begin{equation}
a^\star = \arg\min_{a} \; d\big( \mathbf{s}(\theta^\star), \mathbf{s}(\theta_a) \big), 
\quad \hat{z}^\star = z_{a^\star},
\label{eq:ret}
\end{equation}
where $\theta^\star$ denotes the suspicious model under verification, $\theta_a$ denotes the $a$-th model in the behavior bank $\mathcal{P}$, and $a^\star$ is the index of the most similar reference model. Accordingly, $\hat{z}^\star$ is the inferred provenance label assigned to the suspicious model based on its nearest neighbor. Here, $d(\cdot,\cdot)$ is a distance metric, where we use cosine similarity.

Since different model architectures and training protocols can lead to substantial variations in parameter-level representations, direct comparison at the parameter space becomes unreliable. However, despite these differences, models trained under the same objective with the same dataset tend to learn similar underlying knowledge, which is reflected in their output behaviors. Motivated by this observation, our approach operates in the behavior space, effectively bypassing the instability of parameter-level comparisons. Moreover, this strategy naturally enables black-box verification, as it only requires access to model outputs. In addition, the reference behavior bank can be easily expanded by incorporating more models, thereby further improving robustness and providing strong scalability in practical deployments.

\begin{figure*}
    \centering
    \includegraphics[width=.95\textwidth]{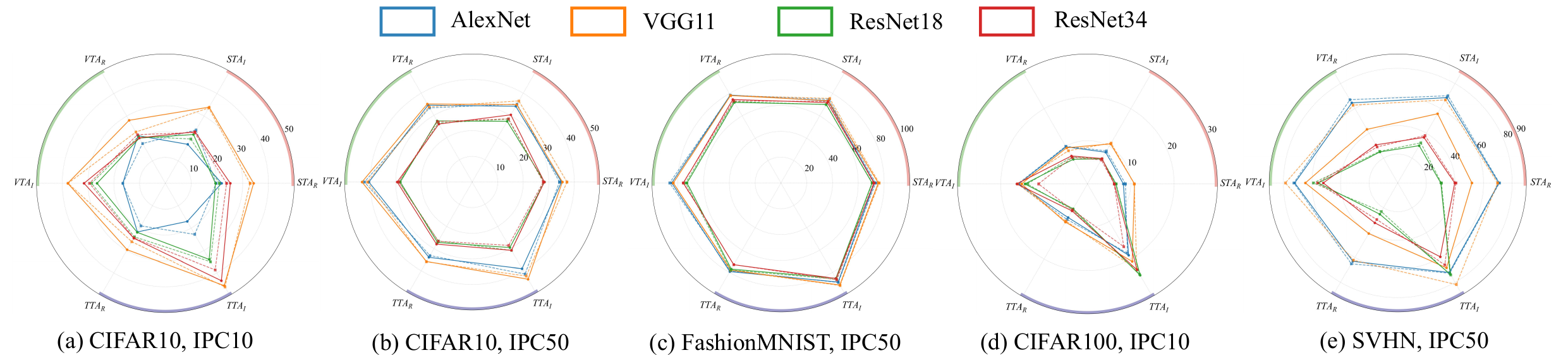}
    \caption{Performance analysis of SubPopMark across diverse datasets and IPC settings using the DC distillation method across different user-side models. Solid lines with circles represent epoch 50, dashed lines with squares represent epoch 100.}
    \vspace{-10pt}
    \label{fig:DC_user1_base_cross}
\end{figure*}

\begin{figure*}[htbp]
    \centering
\includegraphics[width=.95\textwidth]{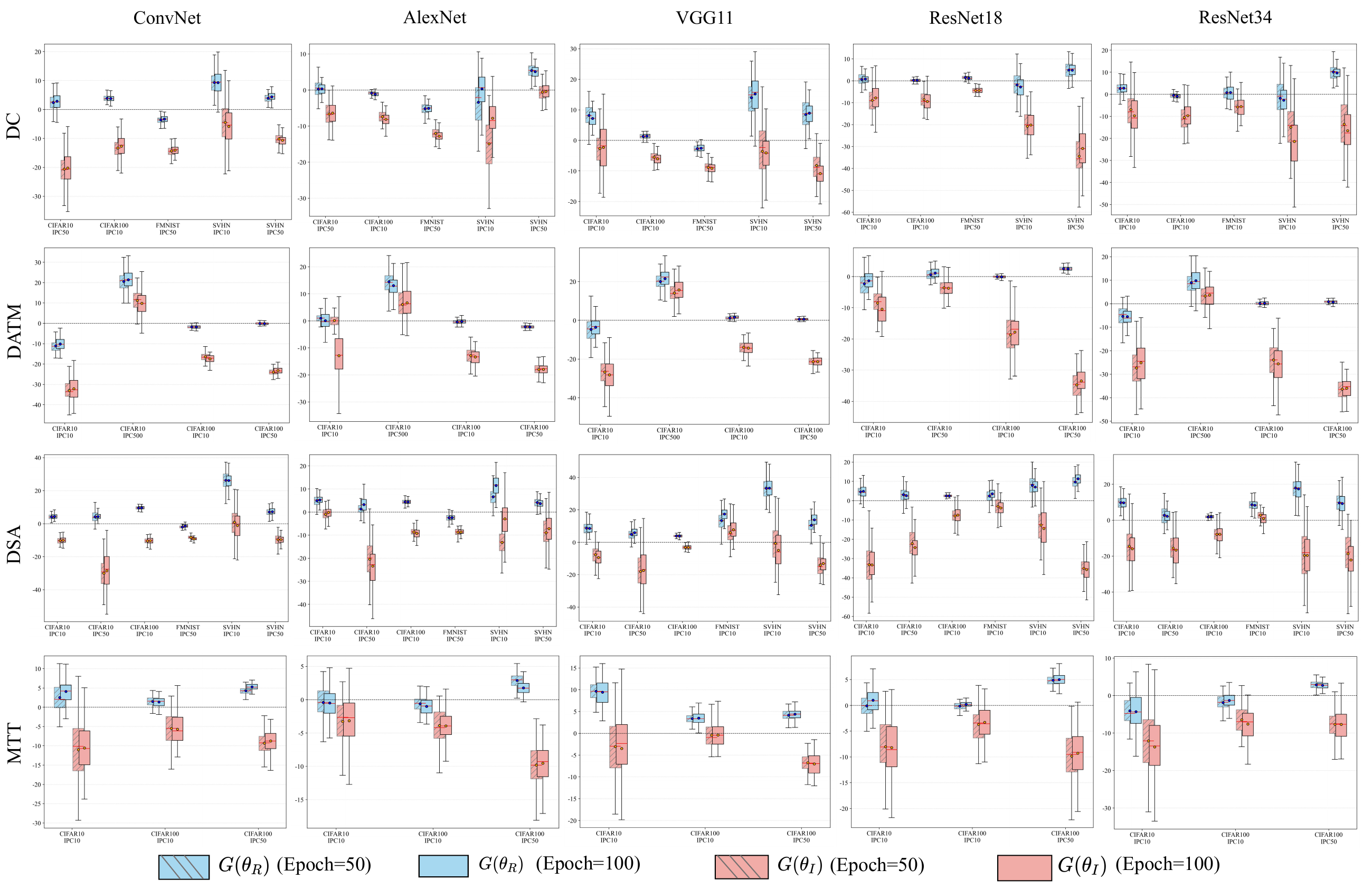}
\vspace{-5pt}
    \caption{Stable reference values of copyright verification performance gaps for reference and infringing models across various architectures and datasets of different DD methods.}
    \vspace{-20pt}
    \label{fig:performance gap}
\end{figure*}

\begin{table}[!t]
\centering
\caption{CVSR (\%) of our method across different DD methods, network architectures, and distilled datasets.}
\label{tab:copyright_verification_rate}
\resizebox{\columnwidth}{!}{%
\begin{tabular}{c|c|c|c|c|c|c|c|c}
\hline
DD method & Dataset & IPC & \diagbox[width=1.6cm]{Epoch}{Arch} 
& ConvNet & AlexNet & VGG11 & ResNet18 & ResNet34 \\
\hline
\multirow{14}{*}{DC} & \multirow{4}{*}{CIFAR10} 
& \multirow{2}{*}{10} & 50  &85.00	&49.00	&76.00	&71.00	&76.00 \\ 
&  &  & 100  &83.00	&67.00	&81.00	&79.00	&88.00 \\  
\cline{3-9} 
&  & \multirow{2}{*}{50} & 50  &98.00	&78.00	&77.00	&60.00	&59.00 \\  
&  &  & 100  &98.00 &76.00 &75.00 &67.00 &56.00 \\  
\cline{2-9} 
& \multirow{2}{*}{CIFAR100} 
& \multirow{2}{*}{10} & 50  &98.50	&98.00	&89.00	&96.00	&93.00 \\ 
&  &  & 100  &98.50	&98.00	&91.00	&94.00	&89.00 \\  
\cline{2-9} 
& \multirow{4}{*}{FashionMNIST} 
& \multirow{2}{*}{10} & 50  &100.00	&54.00	&51.00	&61.00	&63.00 \\ 
&  &  & 100  &97.00	&52.00	&52.00	&57.00	&59.00 \\  
\cline{3-9} 
&  & \multirow{2}{*}{50} & 50  &99.00	&95.00	&64.00	&50.00	&50.00 \\  
&  &  & 100  &100.00	&99.00	&65.00	&50.00	&51.00 \\  
\cline{2-9} 
& \multirow{4}{*}{SVHN} 
& \multirow{2}{*}{10} & 50  &77.00	&64.00	&93.00	&54.00	&60.00 \\ 
&  &  & 100  &84.00	&65.00	&75.00	&86.00	&78.00 \\  
\cline{3-9} 
&  & \multirow{2}{*}{50} & 50  &96.00	&50.00	&81.00	&60.00	&72.00 \\  
&  &  & 100  &94.00 &58.00	&82.00	&62.00	&71.00 \\
\cline{2-9} 
\hline
\hline
\multirow{2}{*}{DM} & \multirow{2}{*}{CIFAR10} 
& \multirow{2}{*}{50} & 50  &99.00	&99.00	&89.00	&81.00	&80.00 \\ 
&  &  & 100  &99.00	&97.00	&88.00	&85.00	&76.00 \\
\cline{2-9} 
\hline
\hline
\multirow{14}{*}{DSA} & \multirow{4}{*}{CIFAR10} 
& \multirow{2}{*}{10} & 50  &94.00	&52.00	&74.00	&89.00	&71.00 \\ 
&  &  & 100  &96.00	&54.00	&72.00	&89.00	&73.00 \\  
\cline{3-9} 
&  & \multirow{2}{*}{50} & 50  &97.50	&90.00	&88.00	&99.00	&91.00 \\  
&  &  & 100 &94.50	&94.00	&86.00	&99.00	&89.00 \\  
\cline{2-9} 
& \multirow{2}{*}{CIFAR100} 
& \multirow{2}{*}{10} & 50  &99.50	&100.00	&90.00	&100.00	&90.00 \\ 
&  &  & 100  &100.00	&100.00	&90.00	&94.00	&90.00 \\  
\cline{2-9} 
& \multirow{4}{*}{FashionMNIST} 
& \multirow{2}{*}{10} & 50  &78.00	&59.00	&53.00	&70.00	&61.00 \\ 
&  &  & 100  &78.50	&60.00	&51.00	&66.00	&59.00 \\  
\cline{3-9} 
&  & \multirow{2}{*}{50} & 50  &97.00	&90.00	&53.00	&50.00	&50.00 \\  
&  &  & 100  &99.50	&94.00	&54.00	&50.00	&50.00 \\  
\cline{2-9} 
& \multirow{4}{*}{SVHN} 
& \multirow{2}{*}{10} & 50  &87.00	&71.00	&89.00	&50.00	&81.00 \\ 
&  &  & 100  &88.00	&83.00	&90.00	&82.00	&96.00 \\  
\cline{3-9} 
&  & \multirow{2}{*}{50} & 50  &100.00	&92.00	&96.00	&100.00	&98.00 \\  
&  &  & 100  &100.00	&92.00	&98.00	&100.00	&98.00 \\
\cline{2-9} 
\hline
\hline

\multirow{12}{*}{DTAM} & \multirow{6}{*}{CIFAR10} 
& \multirow{2}{*}{10} & 50  &95.50&54.00&74.00&28.00&67.00 \\ 
&  &  & 100  &86.50&80.00&94.00&64.00&95.00 \\  
\cline{3-9} 
&  & \multirow{2}{*}{50} & 50 &71.00&62.00&86.00&80.00&67.00 \\  
&  &  & 100 &69.00&72.00&82.00&74.00&71.00 \\  
\cline{3-9} 
&  & \multirow{2}{*}{500} & 50 &83.00&69.00&67.00&52.00&56.00 \\  
&  &  & 100 &79.00&70.00&69.00&52.00&60.00 \\  
\cline{2-9} 
& \multirow{6}{*}{CIFAR100} 
& \multirow{2}{*}{10} & 50  &100.00&99.00&100.00&100.00&99.00 \\ 
&  &  & 100  &100.00&99.00&100.00&99.00&99.00 \\  
\cline{3-9} 
&  & \multirow{2}{*}{50} & 50  &100.00&100.00&100.00&100.00&100.00 \\  
&  &  & 100  &100.00&100.00&100.00&100.00&100.00\\  
\cline{3-9} 
&  & \multirow{2}{*}{100} & 50 &100.00&99.00&92.00&97.00&100.00 \\  
&  &  & 100  &100.00&100.00&94.00&95.00&97.00 \\
\cline{2-9} 
\hline
\hline

\multirow{8}{*}{MTT} & \multirow{4}{*}{CIFAR10} 
& \multirow{2}{*}{10} & 50  &79.50	&68.00	&75.00	&77.00	&56.00 \\ 
&  &  & 100  &83.50	&65.00	&78.00	&72.00	&64.00 \\  
\cline{3-9} 
&  & \multirow{2}{*}{50} & 50  &69.00	&52.00	&53.00	&56.00	&56.00 \\  
&  &  & 100  &64.50	&53.00	&58.00	&62.00	&56.00\\  
\cline{2-9} 
& \multirow{4}{*}{CIFAR100} 
& \multirow{2}{*}{10} & 50  &85.00	&83.00	&63.00	&76.00	&67.00 \\ 
&  &  & 100  &84.50	&83.00	&66.00	&75.00	&74.00 \\  
\cline{3-9} 
&  & \multirow{2}{*}{50} & 50  &97.50	&99.00	&95.00	&95.00	&93.00 \\  
&  &  & 100  &95.50	&98.00	&95.00	&94.00	&93.00 \\ 
\cline{2-9} 
\hline
\end{tabular}%
}
\vspace{-20pt}
\end{table}

\begin{table}[htbp]
\centering
\caption{DLTSR (\%) of the different DD methods across different network architectures and distilled datasets.}
\vspace{-5pt}
\label{tab:dltsr_rate}
\resizebox{\columnwidth}{!}{%
\begin{tabular}{c|c|c|c|c|c|c|c|c}
\hline
DD method & Dataset & IPC & \diagbox[width=1.6cm]{Epoch}{Arch} 
& ConvNet & AlexNet & VGG11 & ResNet18 & ResNet34 \\
\hline
\multirow{14}{*}{DC} & \multirow{4}{*}{CIFAR10} 
& \multirow{2}{*}{10} & 50  &32.00 &9.00 &65.00 &47.00 &37.00 \\ 
&  &  & 100  &38.00 &13.00 &60.00 &43.00 &30.00 \\  
\cline{3-9} 
&  & \multirow{2}{*}{50} & 50  &82.00 &34.00 &65.00 &69.00 &37.00 \\  
&  &  & 100  &92.00 &35.00 &60.00 &57.00 &30.00 \\  
\cline{2-9} 
& \multirow{2}{*}{CIFAR100} 
& \multirow{2}{*}{10} & 50  &100.00 &100.00 &100.00 &100.00 &99.00 \\ 
&  &  & 100  &100.00 &100.00 &100.00 &100.00 &100.00 \\  
\cline{2-9} 
& \multirow{4}{*}{FashionMNIST} 
& \multirow{2}{*}{10} & 50  & 76.00&41.00 &70.00 &88.00 &70.00  \\ 
&  &  & 100  &82.00 &45.00 &73.00 &90.00 &85.00 \\  
\cline{3-9} 
&  & \multirow{2}{*}{50} & 50  &51.00 &40.00 &70.00 &88.00 &70.00 \\  
&  &  & 100  &82.00 &45.00 &73.00 &90.00 &85.00 \\  
\cline{2-9} 
& \multirow{4}{*}{SVHN} 
& \multirow{2}{*}{10} & 50  &64.00 &45.00 &36.00 &83.00 &41.00 \\ 
&  &  & 100  &74.00 &55.00 &45.00 &84.00 &54.00 \\  
\cline{3-9} 
&  & \multirow{2}{*}{50} & 50  &66.00 &69.00 &74.00 &99.00 &79.00 \\  
&  &  & 100  &77.00 &76.00 &78.00 &99.00 &77.00 \\  
\cline{2-9} 
\hline
\hline
\multirow{2}{*}{DM} & \multirow{2}{*}{CIFAR10} 
& \multirow{2}{*}{50} & 50  &93.00	&79.00	&90.00	&94.00	&50.00 \\ 
&  &  & 100  &92.00	&84.00	&94.00	&94.00	&60.00 \\  
\cline{2-9} 
\hline
\hline
\multirow{14}{*}{DSA} & \multirow{4}{*}{CIFAR10} 
& \multirow{2}{*}{10} & 50  &89.00	&40.00	&52.00	&88.00	&53.00 \\ 
&  &  & 100  &89.00	&60.00	&49.00	&88.00	&46.00 \\  
\cline{3-9} 
&  & \multirow{2}{*}{50} & 50  &93.00	&90.00	&98.00	&98.00	&83.00 \\  
&  &  & 100 &94.00	&91.00	&96.00	&95.00	&83.00 \\  
\cline{2-9} 
& \multirow{2}{*}{CIFAR100} 
& \multirow{2}{*}{10} & 50  &100.00	&98.00	&100.00	&100.00	&98.00 \\ 
&  &  & 100  &100.00	&100.00	&100.00	&100.00	&100.00 \\  
\cline{2-9} 
& \multirow{4}{*}{FashionMNIST} 
& \multirow{2}{*}{10} & 50  &56.00	&24.00	&51.00	&87.00	&56.00 \\ 
&  &  & 100  &70.00	&39.00	&54.00	&89.00	&62.00 \\  
\cline{3-9} 
&  & \multirow{2}{*}{50} & 50  &47.00	&20.00	&60.00	&78.00	&70.00 \\  
&  &  & 100  &61.00	&30.00	&59.00	&75.00	&58.00 \\  
\cline{2-9} 
& \multirow{4}{*}{SVHN} 
& \multirow{2}{*}{10} & 50 &85.00	&45.00	&75.00	&97.00	&60.00\\ 
&  &  & 100  &75.00	&68.00	&73.00	&97.00	&70.00\\  
\cline{3-9} 
&  & \multirow{2}{*}{50} & 50  &42.00	&47.00	&65.00	&64.00	&51.00\\  
&  &  & 100  &57.00	&42.00	&62.00	&67.00	&51.00 \\  
\cline{2-9} 
\hline
\hline

\multirow{12}{*}{DTAM} & \multirow{6}{*}{CIFAR10} 
& \multirow{2}{*}{10} & 50  &55.00	&18.00	&71.00	&78.00	&65.00 \\ 
&  &  & 100  &53.00	&40.00	&66.00	&86.00	&60.00 \\  
\cline{3-9} 
&  & \multirow{2}{*}{50} & 50 &46.00	&23.00	&31.00	&45.00	&37.00 \\  
&  &  & 100 &46.00	&35.00	&38.00	&46.00	&38.00 \\  
\cline{3-9} 
&  & \multirow{2}{*}{500} & 50 &50.00	&35.00	&46.00	&47.00	&42.00 \\  
&  &  & 100 &50.00	&39.00	&39.00	&47.00	&40.00 \\  
\cline{2-9} 
& \multirow{6}{*}{CIFAR100} 
& \multirow{2}{*}{10} & 50  &100.00	&100.00	&100.00	&100.00	&100.00 \\ 
&  &  & 100  &100.00	&100.00	&100.00	&100.00	&100.00 \\  
\cline{3-9} 
&  & \multirow{2}{*}{50} & 50  &100.00	&100.00	&100.00	&100.00	&100.00 \\  
&  &  & 100  &100.00	&100.00	&100.00	&100.00	&100.00\\  
\cline{3-9} 
&  & \multirow{2}{*}{100} & 50 &100.00	&100.00	&100.00	&100.00	&100.00 \\  
&  &  & 100  &100.00	&100.00	&100.00	&100.00	&100.00 \\ 
\cline{2-9} 
\hline
\hline

\multirow{8}{*}{MTT} & \multirow{4}{*}{CIFAR10} 
& \multirow{2}{*}{10} & 50  &83.00	&33.00	&63.00	&81.00	&54.00 \\ 
&  &  & 100  &75.00	&26.00	&64.00	&83.00	&50.00 \\  
\cline{3-9} 
&  & \multirow{2}{*}{50} & 50  &47.00	&38.00	&56.00	&62.00	&43.00 \\  
&  &  & 100 &57.00	&42.00	&63.00	&61.00	&43.00\\  
\cline{2-9} 
& \multirow{4}{*}{CIFAR100} 
& \multirow{2}{*}{10} & 50  &100.00	&97.00	&97.00	&100.00	&87.00 \\ 
&  &  & 100 &100.00	&97.00	&100.00	&100.00	&91.00 \\  
\cline{3-9} 
&  & \multirow{2}{*}{50} & 50  &100.00	&100.00	&100.00	&100.00	&96.00\\  
&  &  & 100 &100.00	&96.00	&100.00	&100.00	&100.00 \\  
\cline{2-9} 
\hline
\end{tabular}%
}
\vspace{-15pt}
\end{table}

\begin{table*}[htbp]
    \centering
    \caption{Cross-user leakage tracing on CIFAR-10 (DC, IPC=50). Each entry reports the tracing gap $\hat{G}^i(\theta^j_I)$ (Avg $\pm$ STD, \%). Diagonal entries ($i=j$) indicate correct user attribution, while off-diagonal entries ($i \neq j$) represent cross-user responses.}
    \vspace{-5pt}
    \label{tab:across-user}
    \resizebox{.95\textwidth}{!}{
    \begin{tabular}{c|c|c|c|c|c|c|c|c|c|c|c}
        \toprule
       Infringing Models& \diagbox[width=1.6cm]{Epoch}{$\mathcal{T}_{\text{trac}}$} & $\mathcal{T}^1_{\text{trac}}$ & $\mathcal{T}^2_{\text{trac}}$ & $\mathcal{T}^3_{\text{trac}}$ & $\mathcal{T}^4_{\text{trac}}$ & $\mathcal{T}^5_{\text{trac}}$ & $\mathcal{T}^6_{\text{trac}}$ & $\mathcal{T}^7_{\text{trac}}$ & $\mathcal{T}^8_{\text{trac}}$ & $\mathcal{T}^9_{\text{trac}}$ & $\mathcal{T}^{10}_{\text{trac}}$ \\
        \midrule
        \multirow{2}{*}{$\theta^1_I$} 
        & 50 &\cellcolor{red!20}31.64$\pm$1.43 &18.48$\pm$1.32 &13.52$\pm$0.95 &18.52$\pm$1.57 &20.48$\pm$1.68 &14.72$\pm$1.06 &20.56$\pm$0.70 &20.08$\pm$3.59 &14.60$\pm$0.83 &18.44$\pm$2.31 \\
        & 100 &\cellcolor{red!20}31.68$\pm$1.19 &18.60$\pm$1.49 &13.60$\pm$1.27 &19.16$\pm$1.30 &20.12$\pm$1.49 &16.52$\pm$1.30 &20.16$\pm$0.96 &20.44$\pm$1.04 &13.20$\pm$1.65 &16.84$\pm$1.64 \\
        \midrule
        \multirow{2}{*}{$\theta^2_I$} 
        & 50 &19.89$\pm$1.05 &\cellcolor{red!20}30.26$\pm$1.40 &14.48$\pm$1.32 &16.64$\pm$1.91 &19.48$\pm$1.55 &16.36$\pm$1.02 &20.24$\pm$1.22 &22.60$\pm$0.87 &13.00$\pm$1.26 &17.64$\pm$1.27 \\
        & 100 &20.52$\pm$1.71 &\cellcolor{red!20}31.16$\pm$0.75 &12.16$\pm$0.69  &17.00$\pm$1.25 &20.48$\pm$1.16 &17.00$\pm$1.43 &21.56$\pm$1.30 &21.44$\pm$1.68 &12.80$\pm$0.64 &17.00$\pm$1.52 \\
        \midrule
        \multirow{2}{*}{$\theta^3_I$} 
        & 50 &14.96$\pm$0.34 &14.80$\pm$1.04 &\cellcolor{red!20}43.52$\pm$1.61 &14.48$\pm$0.77 &13.84$\pm$0.37 &16.32$\pm$0.86 &12.04$\pm$0.60 &15.84$\pm$0.48 &14.36$\pm$0.97 &10.20$\pm$0.94 \\
        & 100 &13.08$\pm$1.86 &12.84$\pm$2.03 &\cellcolor{red!20}43.36$\pm$1.49 &14.76$\pm$0.54 &12.40$\pm$1.76 &16.84$\pm$0.57 &12.60$\pm$1.28 &15.24$\pm$1.23 &13.76$\pm$1.85 &10.04$\pm$0.50 \\
        \midrule
        \multirow{2}{*}{$\theta^4_I$} 
        & 50 &17.68$\pm$1.57 &17.57$\pm$0.97 &15.92$\pm$0.89 &\cellcolor{red!20}27.20$\pm$1.56 &18.24$\pm$1.18 &16.28$\pm$1.09 &18.04$\pm$1.21 &18.68$\pm$0.95 &11.28$\pm$0.94 &18.08$\pm$0.80 \\
        & 100 &17.88$\pm$1.03 &16.72$\pm$0.86 &14.80$\pm$0.81 &\cellcolor{red!20}26.68$\pm$1.76 &17.12$\pm$1.24 &16.60$\pm$0.94 &17.96$\pm$1.90 &18.04$\pm$0.60 &10.92$\pm$0.59 &15.88$\pm$0.43 \\
        \midrule
        \multirow{2}{*}{$\theta^5_I$} 
        & 50 &15.28$\pm$1.38 &17.00$\pm$2.32 &9.12$\pm$2.19 &15.08$\pm$0.64 &\cellcolor{red!20}27.60$\pm$2.33 &14.88$\pm$1.43 &17.80$\pm$2.24 &19.28$\pm$0.79 &12.88$\pm$0.99 &15.60$\pm$1.09 \\
        & 100 &16.36$\pm$0.79 &14.80$\pm$1.57 &9.04$\pm$0.82 &15.88$\pm$1.16 &\cellcolor{red!20}29.12$\pm$1.20 &14.36$\pm$0.69 &18.20$\pm$2.19 &18.52$\pm$1.14 &12.08$\pm$1.28 &15.80$\pm$1.88 \\
        \midrule
        \multirow{2}{*}{$\theta^6_I$} 
        & 50 &9.00$\pm$0.95 &11.00$\pm$1.73 &7.00$\pm$0.73 &9.20$\pm$0.90 &11.16$\pm$0.65 &\cellcolor{red!20}28.32$\pm$1.27 &10.56$\pm$0.86 &11.64$\pm$0.41 &11.16$\pm$0.87 &6.32$\pm$2.10 \\
        & 100 &10.20$\pm$1.57 &10.84$\pm$1.50 &7.40$\pm$2.04 &11.28$\pm$1.25 &11.60$\pm$0.77 &\cellcolor{red!20}28.80$\pm$0.46 &9.64$\pm$1.11 &12.80$\pm$1.37 &12.24$\pm$1.70 &5.08$\pm$2.42 \\
        \midrule
        \multirow{2}{*}{$\theta^7_I$} 
        & 50 &19.68$\pm$0.93 &21.92$\pm$1.34 &8.56$\pm$2.45 &16.76$\pm$0.56 &19.52$\pm$0.93 &15.36$\pm$0.94 &\cellcolor{red!20}33.00$\pm$1.54 &21.05$\pm$1.08 &16.16$\pm$1.15 &15.20$\pm$0.77 \\
        & 100 &20.72$\pm$0.63 &22.44$\pm$1.11 &9.12$\pm$1.84 &18.12$\pm$0.74 &21.48$\pm$1.15 &14.36$\pm$1.75 &\cellcolor{red!20}35.48$\pm$0.84 &21.48$\pm$0.92 &15.92$\pm$0.73 &15.76$\pm$1.01 \\
        \midrule
        \multirow{2}{*}{$\theta^8_I$} 
        & 50 &22.16$\pm$0.59 &19.80$\pm$1.56 &13.88$\pm$0.39 &17.68$\pm$1.02 &22.08$\pm$0.41 &17.04$\pm$1.20 &22.20$\pm$1.27 &\cellcolor{red!20}32.00$\pm$0.89 &15.88$\pm$0.93 &20.68$\pm$1.20 \\
        & 100 &21.08$\pm$0.63 &20.44$\pm$1.43 &14.20$\pm$1.05 &17.28$\pm$0.57 &20.64$\pm$0.91 &16.44$\pm$0.80 &22.32$\pm$1.22 &\cellcolor{red!20}31.20$\pm$0.70 &15.28$\pm$0.74 &19.28$\pm$1.09 \\
        \midrule
        \multirow{2}{*}{$\theta^9_I$} 
        & 50 &7.36$\pm$0.41 &8.24$\pm$1.39 &10.56$\pm$1.85 &7.16$\pm$0.85 &12.24$\pm$2.33 &13.80$\pm$1.25 &11.00$\pm$1.76 &11.48$\pm$2.01 &\cellcolor{red!20}21.68$\pm$0.93 &8.60$\pm$2.14 \\
        & 100 &7.88$\pm$1.86 &8.40$\pm$1.79 &8.76$\pm$0.67 &8.60$\pm$1.94 &12.08$\pm$1.00 &13.08$\pm$1.73 &12.60$\pm$2.42 &11.92$\pm$2.42 &\cellcolor{red!20}22.24$\pm$0.65 &8.48$\pm$1.47 \\
        \midrule
        \multirow{2}{*}{$\theta^{10}_I$} 
        & 50 &14.76$\pm$0.74 &16.92$\pm$1.26 &9.28$\pm$1.68 &13.12$\pm$0.47 &15.92$\pm$1.07 &14.96$\pm$0.93 &14.60$\pm$0.67 &18.92$\pm$1.02 &9.28$\pm$0.97 &\cellcolor{red!20}35.28$\pm$1.35 \\
        & 100 &13.48$\pm$0.68 &14.64$\pm$1.53 &9.88$\pm$1.83 &12.08$\pm$1.25 &16.04$\pm$1.71 &13.56$\pm$1.13 &14.16$\pm$1.18 &17.12$\pm$1.38 &9.36$\pm$1.80 &\cellcolor{red!20}34.20$\pm$1.25 \\
        \bottomrule
    \end{tabular}
    }
    \vspace{-15pt}
\end{table*}
\section{Experiment}
\label{sec:experiment}
\subsection{Experimental Setting}

\noindent \textbf{Experiment Environment.} Our proposed method is implemented
using the PyTorch framework and optimized with Stochastic Gradient Descent~(SGD)~\cite{diederik2014adam}. The experiments are conducted on a machine equipped with an AMD Ryzen 7 5800X CPU @3.80 GHz, 32 GB of RAM, and an NVIDIA GTX 3090 GPU.

\noindent \textbf{Experimental setup.} We utilize ConvNet~\cite{krizhevsky2012imagenet} to train the markers. We adpot the ConvNet and AlexNet~\cite{krizhevsky2017imagenet} as the default selection of constructing the reference behavior bank. The user-side models are adopted with ConvNet, AlexNet, VGG11~\cite{simonyan2014very}, ResNet18~\cite{he2016deep},  and ResNet34~\cite{he2016deep}.

\noindent \textbf{DD Methods.} To evaluate the robustness of the proposed method, we conduct experiments under multiple DD methods, including DC~\cite{zhao2021datasetg}, DM~\cite{zhao2023dataset}, DSA~\cite{zhao2021dataset}, DATM~\cite{guo2024lossless}, and MTT~\cite{cazenavette2022dataset}.

\noindent \textbf{Datasets.} To comprehensively evaluate the robustness of the proposed method across varied data distributions, we conduct extensive experiments on multiple standard benchmark datasets, including CIFAR-10~\cite{krizhevsky2009learning}, CIFAR-100~\cite{krizhevsky2009learning}, FashionMNIST~\cite{xiao2017fashion}, and SVHN~\cite{netzer2011reading}.

\noindent \textbf{Metrics.} To comprehensively evaluate the behaivor of the model to different underlying data distributions, we design different evaluation metrics:

\noindent \textbf{\textit{Performance on the Standard Test Set.}} 
We evaluate the model on the standard test set $\mathcal{T}_{\text{stand}}$, drawn from the raw test data distribution, and compute the standard test accuracy~(STA) as $STA = \mathcal{R}(\theta, \mathcal{T}_{\text{stand}})$. This metric reflects the model's performance in the absence of distribution shift.

\noindent \textbf{\textit{Performance on the Copyright Verification Test Set.}}
We further evaluate the model on the copyright verification test set $\mathcal{T}_{\text{veri}}$ constructed via CVM. The verification test accuracy~(VTA) on this set is defined as $VTA = \mathcal{R}(\theta, \mathcal{T}_{\text{veri}})$. This metric measures the model behavior under the verification distribution.

\noindent \textbf{\textit{Performance on the Data Leakage Tracing Test Set.}}
To assess tracing capability, we evaluate the model on the user-specific tracing test set $\mathcal{T}_{\text{trac}}$, and the corresponding tracing test accuracy~(TTA) is defined as $TTA = \mathcal{R}(\theta, \mathcal{T}_{\text{trac}})$, which captures the model performance under the tracing distribution.

Then, to better characterize the behavioral shifts induced by the protected data, we define two performance gap metrics:

\noindent  \textbf{\textit{Copyright Verification Performance Gap:}} $G(\theta) = \mathcal{R}(\theta, \mathcal{T}_{\text{stand}}) - \mathcal{R}(\theta, \mathcal{T}_{\text{veri}})$, which measures the performance drop from the real distribution to the verification distribution, reflecting the model’s sensitivity to CVM-induced distribution shifts.

\noindent  \textbf{\textit{Data Leakage Tracing Performance Gap:}} $\hat{G}(\theta) = \mathcal{R}(\theta, \mathcal{T}_{\text{stand}}) - \mathcal{R}(\theta, \mathcal{T}_{\text{trac}})$, which captures the performance difference between the real distribution and the user-specific tracing distribution, indicating the model’s preference toward a particular user-induced subpopulation.

We further quantify the verification and tracing performance using two metrics:

\noindent \textbf{\textit{Copyright Verification Success Rate (CVSR):}} We propose CVSR to evaluate the effectiveness of our method on the copyright verification task.
\begin{equation}
\text{CVSR} = \frac{1}{M} \sum_{m=1}^{M} 
\mathbb{I}\big({\hat{z}}^\star_m = z_m\big),
\end{equation}
where $\hat{z}^\star_m$ is obtained via our behavior signature retrieval strategy as defined in Eq.~\eqref{eq:ret}, $M$ denotes the total number of evaluated models, and $z_m$ is the ground-truth label.

\noindent \textbf{\textit{Data Leakage Tracing Success Rate (DLTSR):}} We propose DLTSR to evaluate the effectiveness of our method on the data leakage tracing task. Each behavior signature $\mathbf{s}(\theta_a)$ in the reference behavior bank $\mathcal{P} = \{ (\mathbf{s}(\theta_a), z_a) \}_{a=1}^{A}$ is associated with a provenance label $z_a$, which implicitly encodes the user identity $j$. Therefore, the $\hat{z}^\star$ obtained from Eq.~\eqref{eq:ret} can be mapped to a predicted user index ${\hat{j}}^\star$.
\begin{equation}
\text{DLTSR} = \frac{1}{M} \sum_{m=1}^{M} 
\mathbb{I}\big({\hat{j}}^\star = j_m\big),
\end{equation}
where $j_m$ is the corresponding ground-truth label and $M$ represents the total number of evaluated models.

\subsection{Behavioral Analysis}

The core fundation of our method is that models trained on different data distributions exhibit distinguishable behavioral biases toward specific subpopulations. While our theoretical analysis in Sec.~\ref{sec:theory} establishes the emergence of such biases, we further conduct a preliminary empirical study to examine whether they consistently manifest across different models in the DD setting.
Specifically, we first apply our method to protect distilled datasets generated by DC, spanning multiple datasets and diverse distillation settings. To account for variations in downstream training strategies, we further evaluate model performance under different training epochs. The results are shown in Fig.~\ref{fig:DC_base}, we can observe that the STA of the reference model ($STA_R$) remains consistently similar to that of the infringing model ($STA_I$) across all training epochs. It indicates that the introduction of our markers neither causes noticeable performance degradation for the standard test samples, which represents that our method can effectively preserve the knowledge of the distilled dataset. Besides, we observe a noticeable performance gap on $VTA$ and $TTA$ between the reference model ($VTA_R$, $TTA_R$) and the infringing model ($VTA_I$, $TTA_I$). This indicates that our method successfully induces a distribution-specific behavioral bias, which selectively affects the protected subpopulation while preserving performance on standard data. Meanwhile, the performance of the reference model remains consistent across $STA$, $VTA$, and $TTA$, suggesting that our method does not introduce any unintended or malicious behavior into the reference model, and that the observed bias is exclusively induced in the infringing models.

Moreover, we evaluate our method across different downstream models, and the results are shown in Fig.~\ref{fig:DC_user1_base_cross}. Consistent trends can be observed across all models: the performance gap between $VTA$ and $TTA$ remains significant, while the results on $STA$ are relatively stable across all settings. Moreover, we illustrate the copyright verification performance gap between the reference model $G(\theta_R)$ and the infringing model $G(\theta_I)$ across different DD methods and downstream architectures, as shown in Fig.~\ref{fig:performance gap}. It can be observed that the values of $G(\theta_I)$ are mostly below zero, indicating that the infringing models exhibit higher accuracy on the marked data. In contrast, the reference models yield values close to zero, suggesting that they do not exhibit any bias toward the marked samples. This demonstrates that models trained on datasets containing our marker are selectively influenced by the injected subpopulation structure.


\subsection{Copyright Verification Experiment}

Building upon the observed behavioral differences between reference and infringing models, we further investigate whether such differences can be effectively exploited for copyright verification. Specifically, while the previous analysis demonstrates the existence of a systematic performance gap induced by the proposed marker, in this section we evaluate whether this gap can serve as a reliable and robust signal for identifying whether a model has been trained on the protected dataset. 

In this section, we conduct comprehensive experiments across different DD methods and downstream architectures to validate the effectiveness of our copyright verification mechanism. The results in Tab.~\ref{tab:copyright_verification_rate} show that our method achieves consistently high CVSR across a wide range of DD methods, datasets, and downstream architectures. Notably, the verification performance remains generally stable under cross-architecture settings, indicating that the induced prediction bias is largely transferable across different models. Moreover, the method generalizes well across different IPC configurations, with CVSR reaching a high level (e.g., above 80\%) in most cases. These results suggest that the proposed marker provides a robust and reliable prediction bias for copyright verification in the post-distillation setting.

\begin{figure}[!t]
    \centering
    \includegraphics[width=\columnwidth]{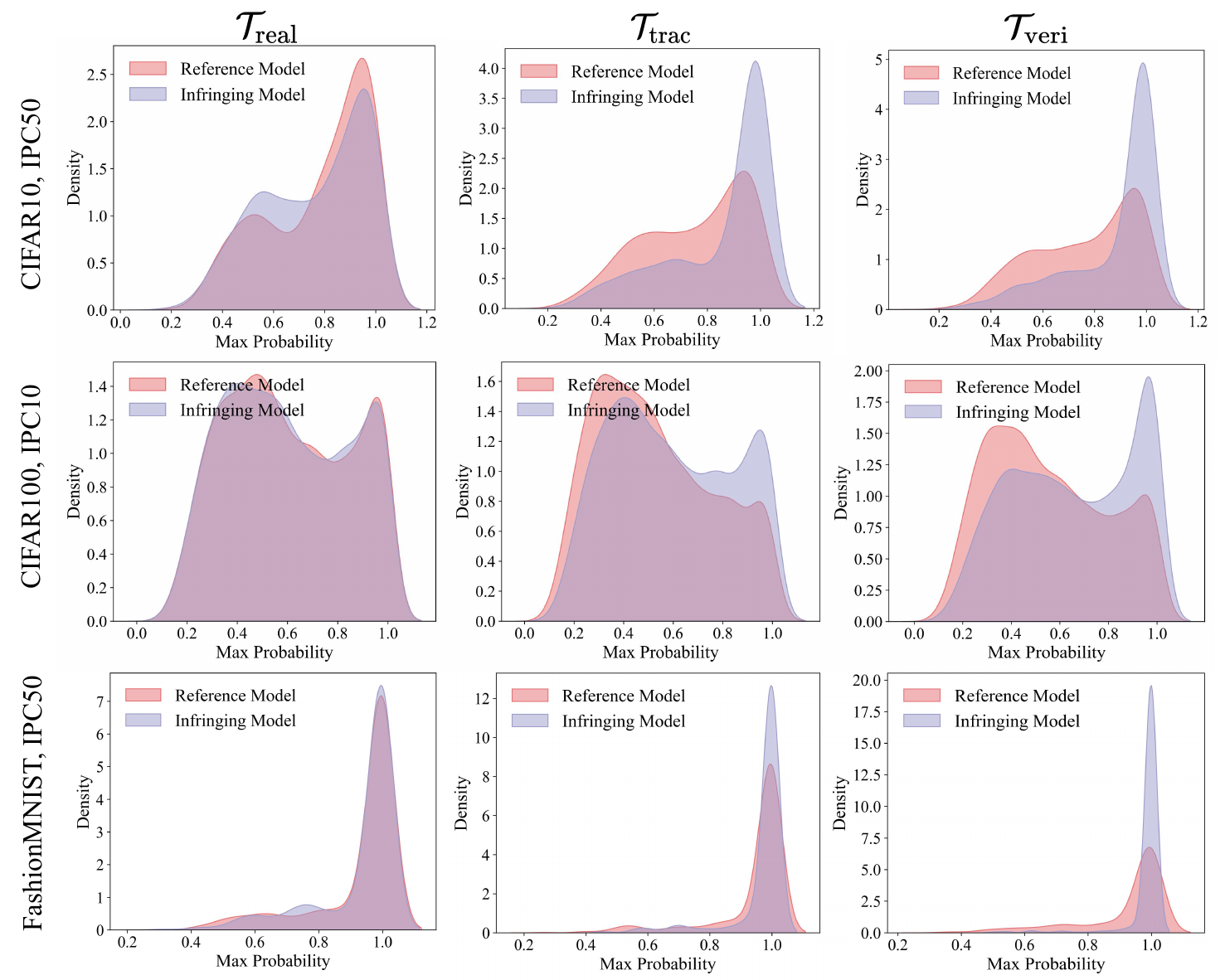}
    \caption{Comparison of prediction distributions between infringing and reference models.}
    \vspace{-15pt}
    \label{fig:DC_distribution}
\end{figure}

\subsection{Data Leakage Tracing Experiment}
Building upon the successful verification of infringing models via CVM, we further investigate whether our method can attribute the source of data leakage to specific users, which aims to answer a more challenging question: which authorized user is responsible for the unauthorized redistribution. To this end, we evaluate the effectiveness of the proposed USTM in inducing a user-specific subpopulation behavior bias. The results are shown in Tab.~\ref{tab:dltsr_rate}. Overall, our method achieves effective data leakage tracing performance across diverse DD methods, datasets, and downstream architectures. These results suggest that the proposed method provides a practical and effective solution for user-level data leakage tracing in the post-distillation setting.

The above experiments demonstrate that our method can effectively trace data leakage at the user level. However, in practical multi-user scenarios, a key challenge arises from the potential conflict between behavior biases induced by different USTM. To investigate this issue, we conduct a cross-user leakage tracing evaluation, where each infringing model is matched against all user-specific tracing sets, and the results are shown in Tab.~\ref{tab:across-user}. It can be observed that, in most cases, the diagonal entries ($i=j$) consistently achieve higher scores than the off-diagonal entries ($i \neq j$), indicating that models respond most strongly to the tracing patterns corresponding to their own subpopulation distribtuion, while exhibiting relatively low sensitivity to the subpopulation distributions associated with other users. 
This can be attributed to the key-conditioned design of USTM. Distinct keys lead to different initialization and optimization trajectories, resulting in diverse subpopulation distributions. Consequently, models learn different behavior biases, responding more strongly to their subpopulation distribution while remaining less sensitive to those of other users.
Therefore, the proposed method supports accurate and stable user-level data leakage tracing even in multi-user settings.
\begin{figure}[!t]
    \centering 
    \includegraphics[width=\linewidth]{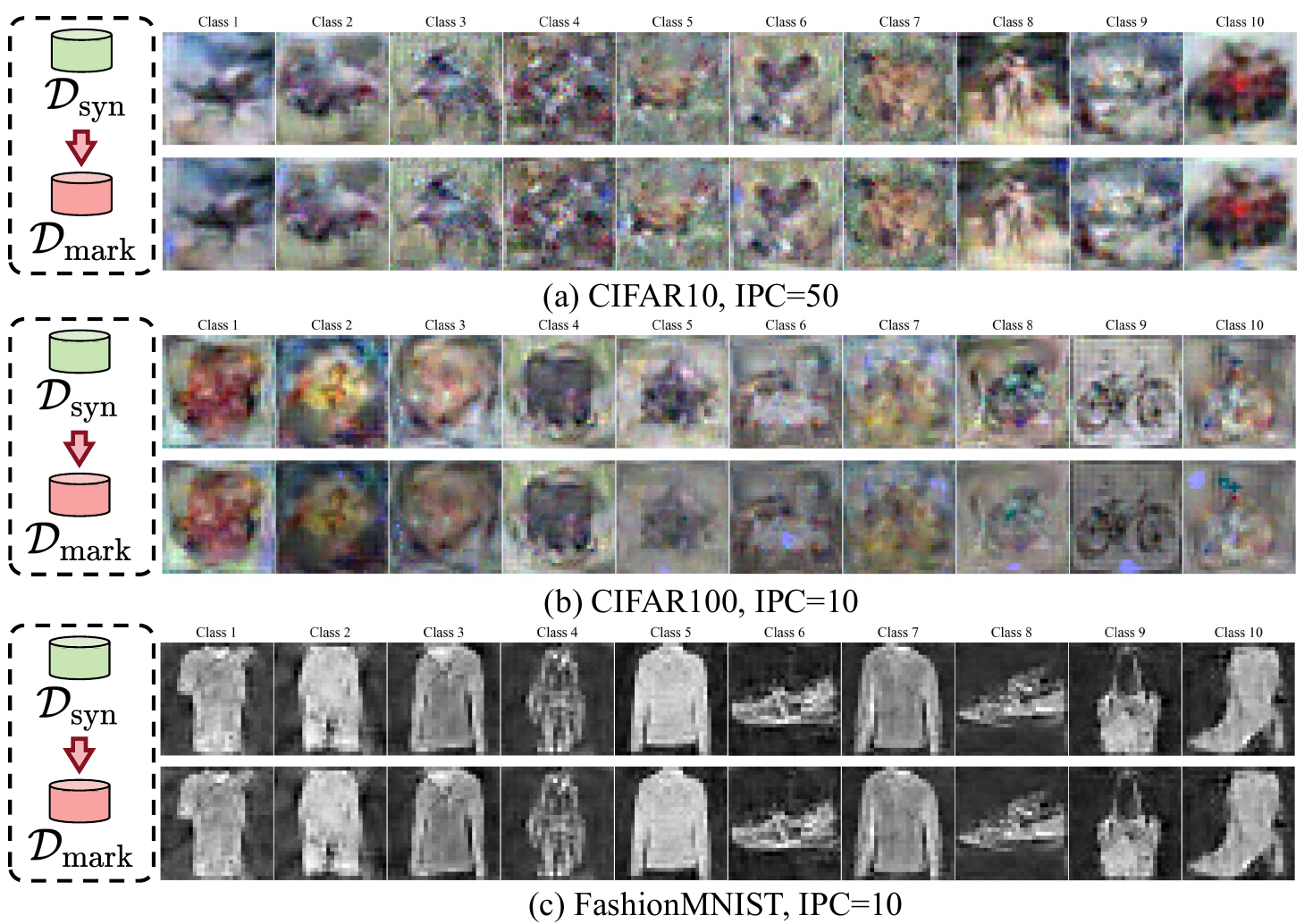}
    \caption{Visualization of the original and our protected distilled datasets.}
    \label{fig:visulize of DC}
    \vspace{-20pt}
\end{figure}

\subsection{Ablation Study}
In this section, we evaluate the effectiveness of different components and hyperparameter choices. The results are shown in Tab.~\ref{tab:psnr_ssim}. It can be observed that our $\mathcal{L}_{\text{per}}$ effectively reduces the perturbation introduced to the distilled dataset, ensuring that the modified samples remain visually consistent with the original ones, as reflected by improved PSNR and SSIM values.
In this section, we evaluate the effectiveness of different components and hyperparameter choices, with results reported in Tab.~\ref{tab:psnr_ssim}. We observe that removing $\mathcal{L}{\text{per}}$ leads to the best CVSR performance, but at the cost of severely degraded visual quality, producing outputs that deviate significantly from the original appearance. In contrast, removing $\mathcal{L}{\text{sem}}$ tends to preserve visual fidelity, but the model fails to form representative subpopulations, limiting its effectiveness. By jointly incorporating both losses, we achieve a favorable balance between reconstruction quality and subpopulation construction.

The results in Tab.~\ref{tab:ablation_accuracy_with_varying_watermark_ratios} demonstrate the impact of different data selection ratios $\alpha$ on model behavior. It can be observed that as $\alpha$ increases, the performance on $STA$ consistently decreases across different datasets, indicating that stronger subpopulation perturbations introduce more pronounced shifts in standard classification behavior for infringing models. In contrast, the gap between $VTA$ and $TTA$ for the infringing and reference models becomes increasingly larger as $\alpha$ increases. This suggests that a larger selection ratio strengthens the induced subpopulation bias, thereby amplifying the behavioral discrepancy between reference and infringing models.

\begin{table}[!t]
\centering
\small
\caption{Ablation study about different losses.}
\label{tab:psnr_ssim}
\resizebox{0.9\linewidth}{!}{
\begin{tabular}{ccccc}
\toprule
Dataset & Method & PSNR (dB) $\uparrow$ & SSIM $\uparrow$ & CVSR (\%) $\uparrow$\\
\midrule
\multirow{3}{*}{\makecell{CIFAR10\\IPC=10}}
& w/o $\mathcal{L}_{\text{per}}$ &10.60 & 0.57 &99.00\\
& w/o $\mathcal{L}_{\text{sem}}$ & 70.69 & 0.99 &57.00\\
& Ours &65.81 & 0.99 &85.00\\
\midrule
\multirow{3}{*}{\makecell{CIFAR10\\IPC=50}}
& w/o $\mathcal{L}_{\text{per}}$ & 8.29 & 0.41 &98.00\\
& w/o $\mathcal{L}_{\text{sem}}$ &69.88 & 0.99 &60.00\\
& Ours &59.26 &0.99 &98.00\\
\midrule
\multirow{3}{*}{\makecell{CIFAR100\\IPC=10}}
& w/o $\mathcal{L}_{\text{per}}$ &9.65 &0.36 &100.00\\
& w/o $\mathcal{L}_{\text{sem}}$ &63.70 &0.99 &66.00\\
& Ours &54.22 &0.99 &98.50 \\
\midrule
\multirow{3}{*}{\makecell{FashionMNIST\\IPC=10}}
& w/o $\mathcal{L}_{\text{per}}$ & 9.35 & 0.61 &98.00\\
& w/o $\mathcal{L}_{\text{sem}}$ &70.84 &0.99 &67.00\\
& Ours & 65.10 &0.99 &100.00\\
\midrule
\multirow{3}{*}{\makecell{FashionMNIST\\IPC=50}}
& w/o $\mathcal{L}_{\text{per}}$ & 7.80 &0.45 & 99.00\\
& w/o $\mathcal{L}_{\text{sem}}$ &68.20 &0.99 &48.00\\
& Ours &61.40 & 0.99 &99.00\\
\midrule
\multirow{3}{*}{\makecell{SVHN\\IPC=10}}
& w/o $\mathcal{L}_{\text{per}}$ & 10.54 & 0.68 &99.00\\
& w/o $\mathcal{L}_{\text{sem}}$ &70.69 &0.99 &53.00\\
& Ours & 57.46 & 0.99 &77.00\\
\midrule
\multirow{3}{*}{\makecell{SVHN\\IPC=50}}
& w/o $\mathcal{L}_{\text{per}}$ & 8.02 &0.45 &98.00\\
& w/o $\mathcal{L}_{\text{sem}}$ &66.21 &0.99 &57.00\\
& Ours & 59.90 &0.99 &96.00\\
\bottomrule
\end{tabular}%
}
\vspace{-15pt}
\end{table}

\subsection{Comparison of Prediction Behavior}
In this section, we further compare the prediction behavior between the reference and the infringing models.
At first, we illustrate the distribution compariosn, and the results are shown in Fig.~\ref{fig:DC_distribution}. It can be observed that the reference and infringing models exhibit similar behavior on $\mathcal{T}_\text{real}$, while showing noticeable differences on the protected datasets $\mathcal{T}_\text{trac}$ and $\mathcal{T}_\text{veri}$. Moreover, the results in Tab.~\ref{tab:ks_p_results} provide a statistical perspective on the behavioral differences between reference and infringing models. Specifically, on the standard test set $\mathcal{T}_{\text{stand}}$, the K-S statistics remain consistently low with large $p$-values, indicating no significant distributional difference between the two models. This suggests that our method preserves the standard generalization performance and does not introduce noticeable bias on natural data. In contrast, on the verification and tracing test sets ($\mathcal{T}_{\text{veri}}$ and $\mathcal{T}_{\text{trac}}$), we observe substantially larger K-S statistics accompanied by extremely small $p$-values (often close to zero), demonstrating statistically significant differences in prediction confidence distributions. This confirms that the proposed markers effectively induce distinguishable behavioral patterns that can be reliably detected.

\subsection{Visualization}
One of our design goals is to ensure that the protection remains imperceptible to human observers, so as to preserve data usability and avoid introducing detectable artifacts that may interfere with downstream applications. We present visualizations of samples from the original distilled dataset and the protected dataset in Fig.~\ref{fig:visulize of DC}. It can be observed that the modifications introduced by our method are visually imperceptible, preserving the overall appearance of the distilled data. This demonstrates that our method does not introduce noticeable artifacts.

\begin{table}[!t]
\caption{Accuracy with varying data selection ratio $\alpha$ of DC method(Avg $\pm$ STD, \%).}
\vspace{-5pt}
\centering
\resizebox{\columnwidth}{!}{
\small
\begin{tabular}{c|c|c|cc|cc|cc}
\hline
& & &\multicolumn{2}{c}{0.1} & \multicolumn{2}{|c}{0.2} & \multicolumn{2}{|c}{0.3} \\ 
\midrule Dataset & IPC & \diagbox[width=1.6cm]{{\small Metric}}{{\small Epoch}} & 50 & 100 & 50 & 100 & 50 & 100 \\
\hline 

\multirow{6}{*}{CIFAR10}
    & \multirow{6}{*}{50} 
    & $STA_R$ 
& 56.52$\pm$1.90
& 57.22$\pm$1.46
  & 56.58$\pm$1.43
  & 56.62$\pm$1.16
  & 55.36$\pm$0.86
  & 57.10$\pm$0.95
    \\
    & & $STA_I$
  & 55.96$\pm$1.37  
  & 55.94$\pm$1.30
  & 48.94$\pm$1.27
  & 48.00$\pm$1.25
  & 44.56$\pm$1.45
  & 45.36$\pm$1.87
    \\ 
    & & $VTA_R$ 
  & 53.96$\pm$2.33 
  & 53.46$\pm$3.48          
  & 56.42$\pm$2.92
  & 57.64$\pm$2.61
  & 53.20$\pm$1.48
  & 54.40$\pm$2.59
    \\ 
    & & $VTA_I$         
  & 79.82$\pm$1.96
  & 78.94$\pm$1.97
  & 95.92$\pm$1.29
  & 97.00$\pm$0.67
  & 98.28$\pm$0.36
  & 98.56$\pm$0.22
    \\              
    & & $TTA_R$            
  & 57.66$\pm$1.78
  & 56.88$\pm$2.40
  & 57.38$\pm$2.76
  & 57.94$\pm$1.43
  & 52.76$\pm$1.45
  & 53.10$\pm$2.08
    \\ 
    & & $TTA_I$          
  & 82.56$\pm$1.36
  & 81.48$\pm$1.52
  & 96.74$\pm$0.98
  & 97.66$\pm$0.58
  & 98.30$\pm$0.30
  & 98.38$\pm$0.41
    \\ 
[0.5ex] \hline \hline

\multirow{6}{*}{CIFAR100} & \multirow{6}{*}{10} 
& $STA_R$ 
  & 24.40$\pm$0.55
  & 24.16$\pm$0.37
  & 24.41$\pm$0.40
  & 24.02$\pm$0.43
  & 24.25$\pm$0.47
  & 24.37$\pm$0.32
     \\
     & & $STA_I$
  & 23.94$\pm$0.43
  & 24.32$\pm$0.47
  & 21.37$\pm$0.38
  & 21.85$\pm$0.37
  & 20.73$\pm$0.52
  & 20.88$\pm$0.43
   \\ 
    & & $VTA_R$          
  & 20.89$\pm$0.94
  & 20.43$\pm$1.11
  & 23.40$\pm$0.79
  & 22.65$\pm$0.55
  & 19.79$\pm$0.97
  & 19.91$\pm$1.14
   \\ 
    & & $VTA_I$         
  & 33.91$\pm$1.07 
  & 34.13$\pm$1.08
  & 49.65$\pm$0.86
  & 50.43$\pm$0.82
  & 69.32$\pm$0.94
  & 69.77$\pm$1.69
   \\ 
    & & $TTA_R$          
  & 22.26$\pm$0.63
  & 21.77$\pm$0.99
  & 16.35$\pm$1.10
  & 15.77$\pm$0.76
  & 19.22$\pm$1.03
  & 19.24$\pm$1.28
   \\ 
    & & $TTA_I$         
  & 45.85$\pm$0.79 
  & 46.12$\pm$0.94
  & 89.95$\pm$0.61
  & 90.59$\pm$1.06
  & 76.39$\pm$0.99
  & 76.82$\pm$1.69
   \\ [0.5ex]
   \cline{2-9}
\hline \hline
\multirow{6}{*}{FashionMNIST} & \multirow{6}{*}{50} 
& $STA_R$
  & 79.00$\pm$1.73
  & 80.94$\pm$0.73
  & 80.88$\pm$1.43
  & 82.72$\pm$0.38
  & 80.32$\pm$1.52
  & 83.14$\pm$0.49
   \\     
    & & $STA_I$            
  & 78.62$\pm$1.45
  & 79.90$\pm$0.90
  & 77.24$\pm$1.49
  & 78.50$\pm$0.45
  & 76.98$\pm$1.20
  & 76.46$\pm$1.64
   \\     
    & & $VTA_R$         
  & 85.22$\pm$2.63
  & 87.48$\pm$0.77
  & 86.78$\pm$1.32
  & 88.22$\pm$0.58
  & 86.70$\pm$1.83
  & 88.86$\pm$0.65
   \\     
    & & $VTA_I$     
  & 93.04$\pm$0.93
  & 94.42$\pm$0.40
  & 96.86$\pm$1.25
  & 97.84$\pm$0.34
  & 97.38$\pm$0.72
  & 97.78$\pm$0.29
   \\                 
    & & $TTA_R$
  & 81.94$\pm$2.79 
  & 83.42$\pm$1.61
  & 79.22$\pm$2.97
  & 82.32$\pm$2.48
  & 81.72$\pm$2.52
  & 84.74$\pm$0.88
   \\     
    & & $TTA_I$         
  & 96.54$\pm$0.65
  & 97.28$\pm$0.20
  & 95.56$\pm$1.14
  & 96.74$\pm$0.28
  & 98.82$\pm$0.17
  & 99.06$\pm$0.25
   \\     
[0.5ex] \hline \hline

\multirow{6}{*}{SVHN} & \multirow{6}{*}{50} 
& $STA_R$ 
  & 81.34$\pm$0.61
  & 81.78$\pm$0.62 
  & 79.54$\pm$0.66
  & 79.60$\pm$0.37
  & 79.54$\pm$0.66
  & 79.60$\pm$0.37
   \\     
    & & $STA_I$            
  & 80.70$\pm$0.45
  & 81.04$\pm$0.45 
  & 78.62$\pm$0.58
  & 78.18$\pm$0.90
  & 76.62$\pm$0.58
  & 77.18$\pm$0.90
   \\     
    & & $VTA_R$     
  & 78.00$\pm$0.98     
  & 77.48$\pm$1.32  
  & 77.64$\pm$1.08
  & 77.44$\pm$1.09
  & 77.64$\pm$1.08
  & 77.44$\pm$1.09
   \\     
    & & $VTA_I$        
  & 77.48$\pm$1.32 
  & 90.74$\pm$0.54  
  & 96.32$\pm$0.47
  & 96.12$\pm$0.44
  & 96.32$\pm$0.47
  & 96.12$\pm$0.44
   \\     
    & & $TTA_R$            
  & 76.04$\pm$1.88
 & 75.76$\pm$2.22
  & 76.78$\pm$1.25
  & 76.52$\pm$0.97
  & 76.78$\pm$1.25
  & 76.52$\pm$0.97
   \\     
    & & $TTA_I$            
  & 93.30$\pm$0.60
  & 94.04$\pm$0.44
  & 97.64$\pm$0.32
  & 97.58$\pm$0.42
  & 97.64$\pm$0.32
  & 97.58$\pm$0.42
   \\     
[0.5ex] \hline
\end{tabular}
\label{tab:ablation_accuracy_with_varying_watermark_ratios}
}
\end{table}

\subsection{Computational Complexity}
\label{sec:comp}
\begin{figure}[!t]
    \centering
    \includegraphics[width=\columnwidth]{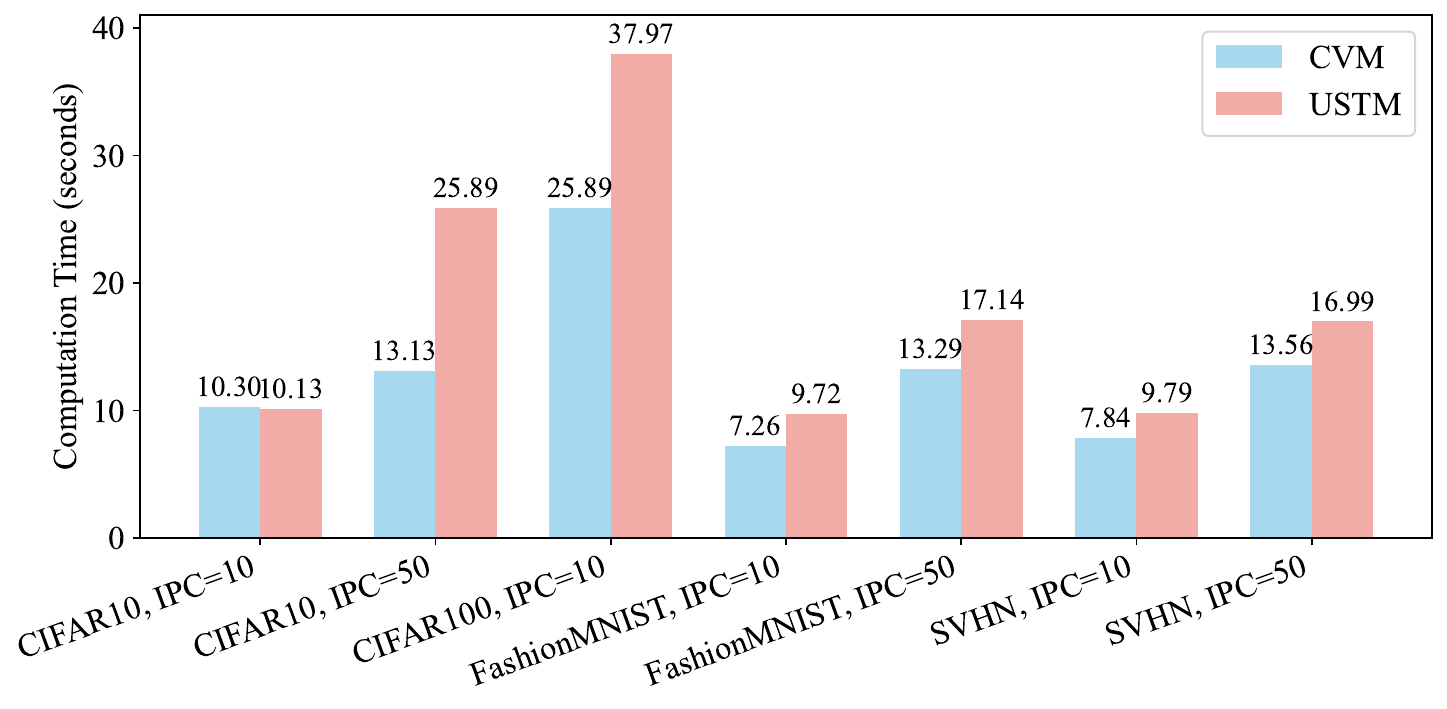}
    \caption{Computational efficiency analysis.}
    \vspace{-15pt}
    \label{fig:DC_times}
\end{figure}

We analyze the computational complexity of our method, and the results are shown in Fig.~\ref{fig:DC_times}. Overall, our proposed protection method is highly efficient, requiring less than one minute to process a distilled dataset in most cases, and in some settings even completing within 10 seconds. This low computational overhead enables fast deployment in practical scenarios, allowing users to seamlessly regenerate protected datasets when switching keys without incurring significant delay.

\begin{table}[!t]
\centering
\small
\caption{Statistical comparison of prediction confidence distributions via the Kolmogorov–Smirnov test, reporting the $d$-statistic and $p$-value between referrence and infringing models.}
\label{tab:ks_p_results}
\resizebox{0.9\linewidth}{!}{
\begin{tabular}{cccc}
\toprule
Dataset & Test set type & \multicolumn{1}{c}{K-S statistic ($d$)} & \multicolumn{1}{c}{$p$-value} \\
\midrule
\multirow{3}{*}{\makecell{CIFAR10\\IPC=10}}
& $\mathcal{T}_{\text{stand}}$ & 0.048 & 0.613 \\
& $\mathcal{T}_{\text{veri}}$ & 0.101 & 1.14e-02 \\
& $\mathcal{T}_{\text{trac}}$ & 0.108 & 5.83e-03 \\
\midrule
\multirow{3}{*}{\makecell{CIFAR10\\IPC=50}}
& $\mathcal{T}_{\text{stand}}$ & 0.062 & 0.060 \\
& $\mathcal{T}_{\text{veri}}$ & 0.332 & 8.54e-25 \\
& $\mathcal{T}_{\text{trac}}$ & 0.360 & 3.52e-29 \\
\midrule
\multirow{3}{*}{\makecell{CIFAR100\\IPC=10}}
& $\mathcal{T}_{\text{stand}}$ & 0.012 & 0.890 \\
& $\mathcal{T}_{\text{veri}}$ & 0.108 & 8.43e-26 \\
& $\mathcal{T}_{\text{trac}}$ & 0.163 & 2.26e-58 \\
\midrule
\multirow{3}{*}{\makecell{FashionMNIST\\IPC=50}}
& $\mathcal{T}_{\text{stand}}$   & 0.026 & 0.996 \\
& $\mathcal{T}_{\text{veri}}$   & 0.172 & 7.10e-07 \\
& $\mathcal{T}_{\text{trac}}$   & 0.356 & 1.57e-28 \\
\midrule
\multirow{3}{*}{\makecell{SVHN\\IPC=50}}
& $\mathcal{T}_{\text{stand}}$ & 0.070 & 0.173 \\
& $\mathcal{T}_{\text{veri}}$ & 0.216 & 1.26e-10 \\
& $\mathcal{T}_{\text{trac}}$ & 0.318 & 9.54e-23 \\
\bottomrule
\end{tabular}%
}
\vspace{-15pt}
\end{table}

\begin{table}[htbp]
\centering
\caption{Performance of our proposed method on Transformer.}
\label{tab:DC_vit}
\resizebox{\columnwidth}{!}{
\begin{tabular}{c|c|c|c|c|c|c|c|c|c}
\hline
Dataset & IPC & Epoch & \diagbox[width=1.6cm]{Metric}{Model} & ConvNet & VGG11 & VIT-Tiny &VIT-Small &CaiT-Small &Swin \\
\hline
\multirow{8}{*}{CIFAR10} 
& \multirow{4}{*}{10} & \multirow{2}{*}{50} & CVSR (\%) &71.50 &67.00 &84.00 &80.00 &94.00 &70.00  \\  
 & & & DLTSR (\%) &32.00 &65.00 &59.00 &70.00 &100.00 &87.00 \\ \cline{3-10} 
 & & \multirow{2}{*}{100} & CVSR (\%) &74.00 &73.00 &84.00 &78.00 &87.00 &60.00 \\  
 & & & DLTSR (\%) &38.00 &60.00 &85.00 &97.00 &100.00 &98.00 \\ \cline{2-10}
& \multirow{4}{*}{50} & \multirow{2}{*}{50} & CVSR (\%) &97.00 &72.00 &78.00 &81.00 &96.00 &70.00 \\  
 & & & DLTSR (\%) &82.00 &65.00 &65.00 &99.00 &100.00 &67.00 \\ \cline{3-10} 
 & & \multirow{2}{*}{100} & CVSR (\%) &97.50 &77.00 &78.00 &77.00 &95.00 &74.00\\  
 & & & DLTSR (\%) &92.00 &60.00 &80.00 &100.00 &100.00 &97.00 \\ 
 [0.5ex] \hline \hline
 \multirow{8}{*}{FashionMNIST} 
& \multirow{4}{*}{10} & \multirow{2}{*}{50} & CVSR (\%) &75.00 &61.00 &59.00 &51.00 &96.00 &64.00 \\  
 & & & DLTSR (\%) &76.00 &70.00 &81.00 &100.00 &99.00 &54.00\\ \cline{3-10} 
 & & \multirow{2}{*}{100} & CVSR (\%) &75.50 &67.00 &64.00 &56.00 &88.00 &66.00\\  
 & & & DLTSR (\%) &82.00 &73.00 &100.00 &100.00 &100.00 &82.00\\ \cline{2-10}
& \multirow{4}{*}{50} & \multirow{2}{*}{50} & CVSR (\%) &99.00 &66.00 &88.00 &88.00 &100.00 &98.00 \\  
 & & & DLTSR (\%)  &51.00 &47.00 &98.00 &100.00 &100.00 &85.00\\ \cline{3-10} 
 & & \multirow{2}{*}{100} & CVSR (\%) &100.00 &65.00 &91.00 &93.00 &99.00 &98.00\\  
 & & & DLTSR (\%) &67.00 &55.00 &97.00 &100.00 &100.00 &90.00 \\ 
 [0.5ex] \hline \hline
  \multirow{8}{*}{SVHN} 
& \multirow{4}{*}{10} & \multirow{2}{*}{50} & CVSR (\%) &77.50 &72.00 &63.00 &52.00 &98.00 &80.00\\  
 & & & DLTSR (\%) &64.00 &36.00 &88.00 &87.00 &100.00 &70.00 \\ \cline{3-10} 
 & & \multirow{2}{*}{100} & CVS (\%)R &83.00 &73.00 &63.00 &55.00 &97.00 &77.00 \\  
 & & & DLTSR (\%) &74.00 &45.00 &98.00 &100.00 &100.00 &95.00 \\ \cline{2-10}
& \multirow{4}{*}{50} & \multirow{2}{*}{50} & CVSR (\%) &99.00 &90.00 &76.00 &51.00 &88.00 &69.00 \\  
 & & & DLTSR (\%) &66.00 &74.00 &75.00 &77.00 &81.00 &53.00 \\ \cline{3-10} 
 & & \multirow{2}{*}{100} & CVSR (\%) &93.00 &91.00 &84.00 &77.00 &90.00 &72.00\\  
 & & & DLTSR (\%) &77.00 &78.00 &80.00 &76.00 &78.00 &74.00 \\ 
 [0.5ex] \hline
\hline
\end{tabular}
}
\end{table}

\subsection{Experiments in Transformers}
All previous experiments are conducted on convolutional architectures. To further evaluate the generality of our method, we extend our experiments to DLTSR-based models. Specifically, building upon the CVM trained on ConvNet, we additionally train CVM and USTM using ViT-Tiny~\cite{dosovitskiy2020image}, resulting in a unified setting where three markers are jointly applied. Correspondingly, the reference behavior bank is also expanded by incorporating ViT-Tiny-based responses. We consider ViT-Small~\cite{dosovitskiy2020image}, CaiT-Small~\cite{Touvron_2021_ICCV}, and Swin Transformer~(Swin)~\cite{liu2021swin} as the user-side models. The results are shown in Tab.~\ref{tab:DC_vit}. It can be observed that our method maintains strong performance on Transformer architectures, demonstrating its generalizability beyond convolutional models. Moreover, even when the Transformer-based marker is combined with the ConvNet-based markers, the proposed framework still preserves effective copyright verification on convolutional architectures, indicating that the protection signals remain robust under heterogeneous backbone settings.



\section{Conclusion}

In this paper, we propose SubPopMark, a harmless framework for copyright protection and data leakage tracing of distilled datasets in a post-distillation setting. Unlike prior work relying on backdoors or label manipulation, our method introduces a carefully designed behavior bias into the models trained on our protected datasets via two components, CVM and USTM, enabling both copyright verification and user-level leakage tracing without affecting semantic labels. We further design a behavior signature retrieval strategy to support downstream-agnostic detection. Extensive experiments across multiple datasets, distillation methods, and architectures demonstrate that SubPopMark achieves strong verification and tracing performance while preserving data utility and visual quality. Overall, this work provides a practical and non-invasive asymmetric protection strategy for distilled datasets. In future work, we will extend this framework to distilled datasets across different modalities.

\bibliography{sample-base}

\end{document}